\documentclass[journal=jpclcd,manuscript=letter]{achemso}
\usepackage[version=3]{mhchem} 
\usepackage[T1]{fontenc}       
\usepackage{amsmath}
\usepackage{braket}
\usepackage{booktabs}
\usepackage{siunitx}
\usepackage{ragged2e}
\usepackage{array}
\usepackage{color}
\usepackage{multirow}
\usepackage{subcaption}
\usepackage{lscape}
\usepackage{hyperref}

\newcolumntype{?}[1]{!{\vrule width #1}}

\SectionNumbersOn
\SectionsOn

\title{Design Principles for Singlet Fission in Aza-BODIPY Dimers: Spacer-Controlled Electronic Structure and Energy Ordering}

\author{Sophiya Goyal}
\affiliation{School of Chemical Sciences and Pharmacy,
	Central University of Rajasthan, Ajmer, India}

\author{S. Rajagopala Reddy}
\email{rajagopala.seelam@curaj.ac.in}
\affiliation{School of Chemical Sciences and Pharmacy,
	Central University of Rajasthan, Ajmer, India}

\begin{document}

\section*{Abstract}
In this work, we investigate the role of spacers in tuning singlet-fission (SF) activity in aza-BODIPY dimers using a diabatic electronic-structure framework. A series of nine spacer-linked dimers is analyzed employing high-level multireference methods to evaluate the energetics and couplings of the locally excited (LE), charge-transfer (CT), and multiexcitonic (ME) states. The introduction of spacers reduces excessive state mixing that was observed in C-C bonded aza-BODIPY dimers and promotes the emergence of more distinct ME states, while preserving the CT-mediated interactions required for efficient SF. Furthermore, the substitution pattern, degree of conjugation, and introduction of acetylene units strongly modulate the relative state energetics and effective couplings, thereby tuning the SF rate over several orders of magnitude. These findings establish clear structure--property relationships and provide molecular design guidelines for optimizing SF activity in aza-BODIPY-based chromophores.

\section{Introduction}
The investigation of triplet-state generation in BODIPY and aza-BODIPY systems has attracted considerable attention over the past decades. Owing to their intrinsic diradical character\cite{Casanova2021} and remarkable photophysical properties,\cite{Wael1977,Killoran2002,Loudet2007} these dyes have emerged as versatile platforms in a wide range of applications, including fluorescent sensing,\cite{Madhu2011,Niu2013} biological imaging,\cite{Ni2014,Kowada2015} and as triplet photosensitizers in solar energy conversion,\cite{Roncali2009,Squeo2020} photodynamic therapy (PDT),\cite{Batat2011,Awuah2012} and photothermal therapy (PTT).\cite{Tian2021,Guo2022} Consequently, significant experimental and theoretical efforts have been devoted to elucidating their excited-state photophysics and underlying mechanisms of triplet-state formation.\cite{Montero2018,Michl2018,Zhao2019,Kandrashkin2024,Casanova2022,Goyal2024}

A combination of advanced spectroscopic techniques, such as nanosecond or femtosecond transient absorption spectroscopy (TAS)\cite{Montero2018} and time-resolved electron paramagnetic resonance (TREPR) spectroscopy,\cite{Zhao2019} together with theoretical approaches including TDDFT\cite{Casanova2022} and CASPT2,\cite{Michl2018} has revealed that the relative torsional orientation between monomer units in covalently linked BODIPY dimers plays a decisive role in dictating the pathway of triplet-state generation. In particular, two competing mechanisms, singlet fission (SF) and spin–orbit charge-transfer intersystem crossing (SOCT-ISC), have been extensively discussed. It has been widely observed that near-orthogonal arrangements between chromophoric units favor the SOCT-ISC pathway, while SF is often considered less favorable due to thermodynamic constraints ($\Delta_{SF} <  0$). However, this picture remains incomplete. The multiconfigurational nature of the ground state in these systems leads to strongly correlated adiabatic excited states, making a simple interpretation based solely on energetic arguments insufficient. As a result, a comprehensive understanding of triplet-state generation requires an explicit treatment of both adiabatic and diabatic representations, enabling a more accurate description of the underlying electronic couplings and competing relaxation pathways.

Building on the need to explicitly account for diabatic states, we previously performed high-level multireference electronic-structure calculations to elucidate their role in triplet-state generation in BODIPY dimers.\cite{Goyal2024} In that work, diabatic electronic states were constructed, the corresponding electronic couplings were evaluated, and SF rate constants ($k_\textrm{SF}$) were estimated. Our analysis of a series of BODIPY dimers revealed consistently low $k_\textrm{SF}$ values at their optimized geometries. This behavior was primarily attributed to the large energy separation between the locally excited (LE) and multiexcitonic (ME) states, which limits efficient population transfer and suppresses the SF efficiency.

Motivated by these findings, we extended our investigation to aza-BODIPY dimers to explore the intramolecular singlet fission (iSF) mechanism in systems with modified electronic structure.\cite{Goyal_arxiv2026} Our results showed that aza-BODIPY dimers exhibit significantly higher $k_\textrm{SF}$ values compared to their BODIPY counterparts, indicating a more favorable scenario for SF. This improvement originates from a reduced energy gap and a more balanced interplay between the relevant diabatic states. In particular, the smaller separation between the LE and charge-transfer (CT) states enhances state mixing and facilitates more efficient coupling pathways.

A detailed diabatic analysis further demonstrated that not only the magnitude of electronic couplings but also the relative energetic ordering of the diabatic states governs the dominant triplet-generation mechanism. Efficient SF requires a favorable energetic alignment in which the ME state lies at or below the LE state, along with sufficient coupling between them. In prototypical systems such as pentacene dimers,\cite{Reddy2022,Reddy2024} this condition is achieved with an energetic ordering of CT $>$ LE $>$ ME, where the CT state remains higher in energy and does not act as a population trap. In contrast, our earlier results for BODIPY dimers showed an unfavorable ordering (ME $>>$ CT $>$ LE), which directs population from LE predominantly toward CT states, thereby promoting the SOCT-ISC pathway over SF.\cite{Goyal2024} For aza-BODIPY dimers, however, a different energetic trend emerges (ME $>$ LE $>$ CT), allowing population transfer from LE into both CT and ME states, depending on the relative coupling strengths. This intermediate regime naturally leads to competition between SF and SOCT-ISC pathways, making these systems particularly suitable for investigating the interplay between the two mechanisms.

Among the C--C bonded regioisomers, the D[1,3] dimer is particularly intriguing because of its asymmetric structure. In our recent study,\cite{Goyal2026} quantum-dynamics simulations revealed coherent SF-like behavior originating from strong mixing among the LE, CT, and ME configurations. None of the adiabatic states possessed a dominant ME character; instead, the relevant excited states were highly mixed and exhibited relatively large diabatic couplings. Such strong state mixing prevents the formation of a well-defined and separated ME state, suggesting that a distinct SF pathway is unlikely in this dimer and instead indicating a greater probability of SOCT-ISC. To further investigate this possibility, intersystem crossing rates and spin--orbit couplings were analyzed, revealing that SOCT-ISC can occur from $\textrm{S}_{1}$ to $\textrm{T}_{3}$ in this system. Similar behavior was also observed for the other regioisomers, D[1,1], D[3,3], and D[2,2], where strong electronic-state mixing likewise prevented the isolation of a distinct ME state.\cite{Goyal_arxiv2026} The absence of an efficient SF pathway in these dimers can therefore be attributed to the strong mixing of the adiabatic states arising from the short intermonomer distance.
Based on these observations, we hypothesized that introducing spacers between the monomer units would increase the intermonomer separation, allow a more distinct ME state to emerge, and modify the diabatic energetic ordering particularly by pushing the CT state to higher energies, thereby increasing the likelihood of efficient SF.

\begin{figure}[!ht]
	\centering
	\includegraphics[width=0.8\textwidth]{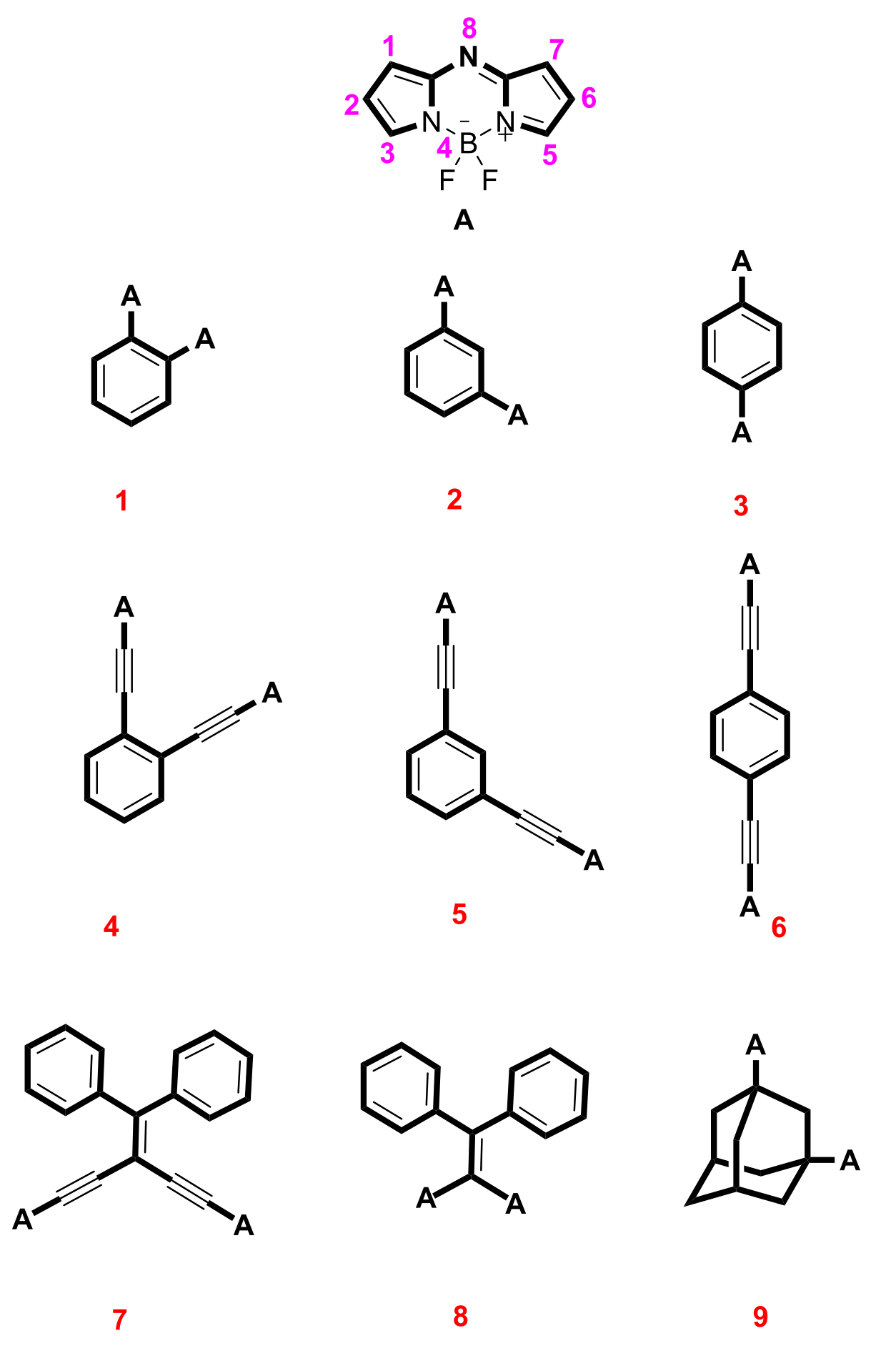}
	\caption{Graphical representations of the spacer-linked aza-BODIPY dimers (D[3,3] series) investigated in this work.}
	\label{fig:aza-spacer} 
\end{figure}

To examine this idea, we perform electronic-structure calculations and evaluate the corresponding diabatic electronic couplings for a series of spacer-linked systems (Figure~\ref{fig:aza-spacer}). The selected spacers enable us to examine the effects of conjugation, resonance, and structural connectivity on the SF-relevant diabatic states and their electronic interactions. In total, nine dimers are considered, including benzene-spacer-linked systems connected through the ortho, meta, and para positions, both with and without an acetylene unit (systems 1--6). Recent studies have examined the role of the acetylene unit in pentacene dimers in the context of iSF.\cite{Majumder2024} Specifically, 2,2$^{\prime}$- and 6,6$^{\prime}$-linked dimers were investigated with and without the acetylene unit. The results showed that the presence of the acetylene unit significantly influences the iSF process. Among the systems studied, the 2Ac-P (acetylene-bridged 2,2$^{\prime}$) dimer exhibited the slowest iSF dynamics, which was attributed to its larger intermonomer distance and weaker electronic coupling. Motivated by this, the effect of the acetylene unit will be further evaluated here through comparative electronic-structure calculations on representative dimer pairs with and without the acetylene unit.

To further analyze the role of conjugation, the systems are classified into three categories: fully conjugated benzene-bridged dimers, a cross-conjugated dimer (system 7), and a non-conjugated adamantane-linked dimer (system 9). For all systems, we characterize the adiabatic states and evaluate the diabatic electronic couplings relevant to SF. This analysis allows us to examine how spacer topology and conjugation influence the energetic alignment and coupling between the LE, CT, and ME states.

\section{Methods}
\label{sec:method}
The ground-state geometries were optimized using restricted Becke’s three-parameter hybrid exchange functional combined with the Lee--Yang--Parr correlation functional (RB3LYP)\cite{Becke1993,Stephens1994} in conjunction with Dunning’s correlation-consistent polarized valence double-zeta (cc-pVDZ) basis set,\cite{Jr1989} as implemented in Gaussian 09.\cite{Frisch2009}  Owing to the multiconfigurational nature of the ground electronic state, the electronic structure of these systems requires multireference methods for a reliable description. Therefore, multireference excited-state calculations were carried out using the state-averaged extended multiconfiguration quasi-degenerate perturbation theory\cite{Granovsky2011} (SA15-XMCQDPT[8,8]) method with the cc-pVDZ basis set\cite{Dunning2013} for singlet, triplet, and quintet states, as implemented in the GAMESS package.\cite{Gordon2020} Here, SA15 denotes state averaging over 15 electronic states, and the (8,8) active space corresponds to 8 electrons distributed in 8 active orbitals. A shift of 0.04 was applied to reduce the intruder-state effect. Further analysis was carried out to understand the SF mechanism in these dimers. The general mechanism of SF in related systems was discussed in our previous study.\cite{Goyal2024}

To gain deeper insight into the iSF mechanism, the couplings between the relevant diabatic electronic states were evaluated\cite{Zirzlmeier2016,Basel2017,Reddy2018,Reddy2019} using the fourfold diabatization approach of Truhlar and Nakamura.\cite{Nakamura2001,Nakamura2002} In total, eight diabatic states relevant to SF were considered: the ground state $^1\left(\textrm{S}_\textrm{0} \textrm{S}_\textrm{0}\right)$, one ME state $^1\left(\textrm{T}_\textrm{1} \textrm{T}_\textrm{1}\right)$, two LE states in which one monomer is in its ground state and the other is in its first excited state ($^1\left(\textrm{S}_\textrm{1} \textrm{S}_\textrm{0}\right)$ and $^1\left(\textrm{S}_\textrm{0} \textrm{S}_\textrm{1}\right)$), two CT configurations in which an electron from the HOMO of one monomer is transferred to the LUMO of the other monomer ($^{1}(\textrm{C}\textrm{A})$ and $^{1}(\textrm{A}\textrm{C})$), where C and A denote the radical cation and radical anion forms of the BODIPY moiety, respectively, and two double-excitation (DE) states $^{1}\left(\textrm{DE}_\textrm{1}\right)$ and $^{1}\left(\textrm{DE}_\textrm{2}\right)$.

The diabatization procedure included all configuration state functions (CSFs) with coefficients greater than 0.20 across the adiabatic electronic states. After applying the three-fold density criterion and the MORMO conditions, the diabatic molecular orbitals (DMOs) were derived from the adiabatic (canonical) molecular orbitals. Finally, a unitary transformation was used to convert the adiabatic states represented in the DMO basis into diabatic states. The relative contributions of the dominant CSFs derived from the DMOs to the adiabatic states S$_0$, S$_1$, S$_2$, S$_3$, S$_4$, S$_5$, S$_6$, and S$_7$ for all geometries were approximately 94\% or higher.

The effective coupling $V_\text{eff}$ between these diabatic states was computed using the following expression. \cite{Berkelbach2013,Reddy2018a}
\begin{equation}
	\begin{split}
		V_{eff}  & \approx \bra{^{1}(\textrm{S}_{1}\textrm{S}_{0})}V\ket{^{1}(\textrm{T}_{1}\textrm{T}_{1})}\\
		&-2\tfrac{\bra{^{1}(\textrm{S}_{1}\textrm{S}_{0})}V\ket{^{1}(\textrm{CA})}\bra{^{1}(\textrm{CA)}}V\ket{^{1}(\textrm{T}_{1}\textrm{T}_{1})}+\bra{^{1}(\textrm{S}_{1}\textrm{S}_{0})}V\ket{^{1}(\textrm{AC})}\bra{^{1}(\textrm{AC}}V\ket{^{1}(\textrm{T}_{1}\textrm{T}_{1})}}{[E({^{1}(\textrm{CT})})- E({^{1}(\textrm{T}_{1}\textrm{T}_{1})}) + E({^{1}(\textrm{CT})})- E({^{1}(\textrm{S}_{1}\textrm{S}_{0})})]}.
		\label{eq:eff_coup}
	\end{split}
\end{equation}
\noindent In this equation, $\bra{\Psi}V\ket{\Phi}$ denotes the coupling between the diabatic states $\Psi$ and $\Phi$. The terms $E(^{1}(\textrm{CT}))$, $E(^{1}(\textrm{T}_\textrm{1}\textrm{T}_\textrm{1}))$, $E(^{1}(\textrm{S}_\textrm{1}\textrm{S}_\textrm{0}))$ reflect the  energies associated with the diabatic states $^{1}(\textrm{CT})$, $^{1}(\textrm{T}_\textrm{1}\textrm{T}_\textrm{1})$  and $^{1}(\textrm{S}_\textrm{1}\textrm{S}_\textrm{0})$ respectively, evaluated at the geometry optimized in the ground state. The first term in the expression corresponds to the direct mechanism, and the second term describes the mediated mechanism. The rate constant ($k_\text{SF}$) for LE to ME state conversion, is calculated using:
\begin{equation}
	k_\textrm{SF} = \frac{2 \pi}{\sqrt{4 \pi \hbar^2 \lambda k_B T}}  \vert V_{eff} \vert ^2  \exp{\left ({-\frac{[E({^{1}(\textrm{T}_\textrm{1}\textrm{T}_\textrm{1})})-(E({^{1}(\textrm{S}_\textrm{1}\textrm{S}_\textrm{0})})+\lambda)]^2}{4\lambda k_{B} T}}\right )}
	\label{eq:sf_rate}
	\vspace{0.1cm}
\end{equation}
\noindent Here, $\lambda$ represents the reorganization energy.\cite{Nitzan2006} Using $V_\text{eff}$ and assigning $\lambda$ a value of 100 meV allowed the
determination of the rate constant $k_\text{SF}$ in order to generate the ME state within the studied dimers.

\section{Results}
\subsection{Vertical Excited States}

Tables \ref{tab:excited_states1234}–\ref{tab:excited_states8910} summarize the computed vertical excitation energies (VEEs, in eV), oscillator strengths ($f$), dipole moments ($\mu$, in Debye), and the corresponding state characterizations for all dimers studied. For each system, the eight lowest singlet states, three lowest triplet states, and the lowest quintet state are reported. Because the optimized molecular orbitals of these dimers are intrinsically delocalized, the excited-state analysis was carried out using a (4,4) active space constructed from localized orbitals. Inspection of the lowest excited states reveals several distinct classes of excitations. Two types of LE states are identified:(1) a HOMO→LUMO excitation, denoted LE$_\textrm{1}$, and (2) a HOMO–1→LUMO excitation, referred to as LE$_\textrm{2}$. Similarly, two CT configurations are observed: (1) CT$_\textrm{1}$, corresponding to HOMO(monomer A)→LUMO(monomer B), and (2) CT$_\textrm{2}$, arising from HOMO–1(monomer A)→LUMO(monomer B).

\begin{landscape}
	\begin{table}[!ht]
		\centering
		\caption[The VEEs, oscillator strengths, dipole moments and character of aza-BODIPY dimers 1, 2 and 3.]{The vertical excitation energies (VEEs in eV units)$^a$, oscillator strengths ($f$, in dimensionless units)$^b$, dipole moments ($\mu$, in Debye units)$^b$, and character$^c$ of the electronic state of aza-BODIPY dimers 1, 2 and 3.} 
		\begin{tabular}{rrrrrrrrrrrrrrrr}
			\toprule
			\multirow{2}{*}{State} && \multicolumn{4}{c}{1} & & \multicolumn{4}{c}{2} & & \multicolumn{4}{c}{3} \\
			\cmidrule[0.5pt]{3-6} \cmidrule[0.5pt]{8-11} \cmidrule[0.5pt]{13-16}
			&& VEEs & $f$ & $\mu$ &char &&  VEEs & $f$ & $\mu$ &char && VEEs & $f$ & $\mu$  &char \\
			\toprule
			S$_\text{0}$ && 0.000 & 0.000&2.45  & GS &&0.000 & 0.000& 3.43& GS&& 0.000& 0.000& 0.30& GS\\
			S$_\text{1}$ && 1.950 & 1.729&0.94 & LE$_1$&& 2.081 & 2.089& 3.77& LE$_1$&& 1.976 & 2.054& 1.49 & LE$_1$ \\
			S$_\text{2}$ && 2.257& $<$0.001&2.16 & ME&& 2.322 & 0.393& 3.94& LE$_1$&& 2.170 & $<$0.001 & 0.21  & ME  \\
			S$_\text{3}$ && 2.324& 0.095&0.15 & LE$_1$&& 2.447 & $<$0.001& 2.67& ME&& 2.354 & 0.444 & 1.97 &  LE$_1$ \\
			S$_\text{4}$ && 3.066& $<$0.001&2.11 & CT$_1$&& 2.943 & 0.074& 2.13& CT$_1$&& 3.178 & 0.008 & 1.07 &  CT$_1$ \\
			S$_\text{5}$ && 3.125& 0.115&1.39 & CT$_1$&& 3.060 & 0.251& 2.53& CT$_1$&& 3.329 & 0.170 & 1.97 &  CT$_1$\\
			S$_\text{6}$ && 3.634& 0.122&1.53 & LE$_2$&& 3.665 & 0.007& 2.30& DE$_1$&& 3.409 & 0.003& 0.54 & LE$_2$  \\
			S$_\text{7}$ && 3.904& 0.037&1.65 & LE$_2$&& 3.746 & 0.003& 2.77& DE$_2$&& 3.722 & 0.006& 0.30 & LE$_2$  \\
			T$_\text{1}$ && 1.163& 0.000& 2.40& $-$&& 1.153 & 0.000& 2.74& $-$&& 1.112 & 0.000& 2.74 &$-$   \\
			T$_\text{2}$ && 1.249& 0.000& 2.34& $-$&& 1.174 & 0.000& 2.72& $-$&& 1.241 & 0.000& 2.72 & $-$ \\
			T$_\text{3}$ && 2.380& 0.000& 2.30& $-$&& 2.455 & 0.000& 2.39& $-$&& 2.287 & 0.000& 2.39 & $-$ \\
			Q$_\text{1}$ && 2.684& 0.000& 2.10& $-$&& 2.521 & 0.000& 2.72 & $-$&& 2.639 &0.000 & 2.72 &$-$   \\
			\bottomrule		
			\label{tab:excited_states1234}
		\end{tabular}\\
	\small $^a$ Calculated at SA15-XMCQDPT(8,8)/cc-pVDZ level of theory.\\ $^b$ Calculated at SA15-CASSCF(8,8)/cc-pVDZ level of theory. \\ $^c$ Character of the excited states: GS denotes the ground state. LE$_1$ and LE$_2$ are optically allowed states arising from symmetric and antisymmetric combinations of local excitations within each aza-BODIPY monomer, corresponding to  HOMO→LUMO and  HOMO-1→LUMO transitions, respectively. ME denotes the multiexcitonic state, and CT$_1$ represents charge-transfer states involving electron transfer from the HOMO of one monomer to the LUMO of the other. \\ 
	\end{table}
\end{landscape}

\begin{landscape}
	\begin{table}[!ht]
		\centering
		\caption[The VEEs, oscillator strengths, dipole moments and character of aza-BODIPY dimers 4, 5 and 6.]{The vertical excitation energies (VEEs in eV units)$^a$, oscillator strengths ($f$, in dimensionless units)$^b$, dipole moments ($\mu$, in Debye units)$^b$, and character$^c$ of the electronic state of aza-BODIPY dimers 4, 5 and 6.}
		\begin{tabular}{rrrrrrrrrrrrrrrr}
			\toprule
			\multirow{2}{*}{State} && \multicolumn{4}{c}{4} & & \multicolumn{4}{c}{5} & & \multicolumn{4}{c}{6} \\
			\cmidrule[0.5pt]{3-6} \cmidrule[0.5pt]{8-11} \cmidrule[0.5pt]{13-16} 
			&& VEEs & $f$ & $\mu$ &char &&  VEEs & $f$ & $\mu$ &char &&  VEEs & $f$ & $\mu$ &char \\
			\toprule
			S$_\text{0}$ && 0.000 & 0.000& 2.43& GS&&0.000 & 0.000& 1.01 & GS&&0.000 & 0.000& 1.98& GS \\
			S$_\text{1}$ && 2.069 & 2.463& 2.14& LE$_1$&&2.053 & 2.214& 1.61 & LE$_1$&& 1.945  & 2.546 & 2.13& LE$_1$ \\
			S$_\text{2}$ && 2.242 & $<$0.001& 2.19& ME&&2.243 & 0.521 & 1.62 & LE$_1$&& 2.155  & $<$0.001 & 1.94 & ME  \\
			S$_\text{3}$ && 2.315 & 0.024& 2.73& LE$_1$&&2.294 &  $<$0.001& 0.68 & ME&& 2.228  & 0.012 & 2.02 &  LE$_1$  \\
			S$_\text{4}$ && 2.983 & 0.002& 2.09& CT$_1$&&2.957 & 0.147 & 0.03 & CT$_1$&& 2.998  & $<$0.001 & 1.98 &  CT$_1$ \\
			S$_\text{5}$ && 3.172 & 0.198& 3.80& CT$_1$&&3.281 & 0.094 & 0.69 & CT$_1$&& 3.113  & 0.217 & 1.90 & CT$_1$  \\
			S$_\text{6}$ && 3.456 & 0.123& 2.52& DE$_1$&&3.515 & 0.002 & 0.38 & DE$_1$&& 3.278  & 0.453 & 2.10 &  DE$_1$ \\
			S$_\text{7}$ && 3.563 & 0.009& 2.33& DE$_2$&&3.639 & $<$0.001& 0.57 & DE$_2$&& 3.594  & $<$0.001& 2.11 &    DE$_2$ \\
			T$_\text{1}$ && 1.093 & 0.000& 2.17& $-$&&1.078 & 0.000& 0.51 &$-$ && 1.057  & 0.000& 1.93 & $-$   \\
			T$_\text{2}$ && 1.125 & 0.000& 2.16& $-$&&1.090 & 0.000& 0.50 & $-$&& 1.126  & 0.000& 1.93 & $-$   \\
			T$_\text{3}$ && 2.258 & 0.000& 2.07& $-$&&2.304 & 0.000& 0.25 & $-$&& 2.211  & 0.000& 1.90 & $-$   \\
			Q$_\text{1}$ && 2.493 & 0.000& 2.07& $-$&&2.212 & 0.000& 0.70 & $-$&& 2.511  & 0.000& 1.91 & $-$   \\
			\bottomrule
			\label{tab:excited_states567}
		\end{tabular}\\
		\small $^a$ Calculated at SA15-XMCQDPT(8,8)/cc-pVDZ level of theory.\\ $^b$ Calculated at SA15-CASSCF(8,8)/cc-pVDZ level of theory. \\ $^c$ Character of the excited states: GS denotes the ground state. LE$_1$ and LE$_2$ are optically allowed states arising from symmetric and antisymmetric combinations of local excitations within each aza-BODIPY monomer, corresponding to  HOMO→LUMO and  HOMO-1→LUMO transitions, respectively. ME denotes the multiexcitonic state, and CT$_1$ represents charge-transfer states involving electron transfer from the HOMO of one monomer to the LUMO of the other.
	\end{table}
\end{landscape}

\begin{landscape}
	\begin{table}[!ht]
		\centering
		\caption[The VEEs, oscillator strengths, dipole moments and character of aza-BODIPY dimers 7, 8 and 9.]{The vertical excitation energies (VEEs in eV units)$^a$, oscillator strengths ($f$, in dimensionless units)$^b$, dipole moments ($\mu$, in Debye units)$^b$, and character$^c$ of the electronic state of aza-BODIPY dimers 7, 8 and 9.}
		\begin{tabular}{rrrrrrrrrrrrrrrr}
			\toprule
			\multirow{2}{*}{State} && \multicolumn{4}{c}{7} & & \multicolumn{4}{c}{8} & & \multicolumn{4}{c}{9}\\
			\cmidrule[0.5pt]{3-6} \cmidrule[0.5pt]{8-11} \cmidrule[0.5pt]{13-16}
			&& VEEs & $f$ & $\mu$ &char &&  VEEs & $f$ & $\mu$ &char && VEEs & $f$ & $\mu$ &char \\
			\toprule
			S$_\text{0}$ && 0.000  & 0.000& 2.16& GS&& 0.000& 0.000& 88.85& GS&& 0.000 & 0.000&  15.59&  GS\\
			S$_\text{1}$ && 1.929  & 0.577& 4.81& LE$_1$&& 1.902& 0.896& 81.93& LE$_1$&&2.214 & 1.002&  16.24& LE$_1$ \\
			S$_\text{2}$ && 1.978  & 1.951& 6.95& LE$_1$&& 2.256& 0.016& 88.03& LE$_1$&&2.259 & $<$0.001& 9.23 & LE$_1$\\
			S$_\text{3}$ && 2.322  & $<$0.001& 2.08& ME&& 2.316& 0.033& 77.45& ME&&2.423 & 0.137&  51.48&  ME\\
			S$_\text{4}$ && 2.406  & 0.078& 3.89& CT$_1$&& 2.971& 0.339& 77.21& CT$_1$&&3.148 & $<$0.001&  31.90& LE$_2$ \\
			S$_\text{5}$ && 2.583  & 0.108& 3.70& LE$_2$&& 3.121& 0.747& 89.46& CT$_1$&&3.159 & 0.083&  41.45& LE$_2$ \\
			S$_\text{6}$ && 2.921  & 0.384& 3.12& CT$_1$&& 3.248& 0.229& 78.93& LE$_2$&&4.116 & $<$0.001& 67.14 & CT$_1$ \\
			S$_\text{7}$ && 3.206  & 0.035& 1.67& LE$_2$&& 3.626& 0.097& 75.22& LE$_2$&&4.122 & 0.142& 27.34 &  CT$_1$\\
			T$_\text{1}$ && 0.811  & 0.000& 1.92& $-$&& 1.293& 0.000& 76.43& $-$&&1.247 & 0.000&  54.38& $-$ \\
			T$_\text{2}$ && 0.850  & 0.000& 1.82& $-$&& 1.370& 0.000& 77.58& $-$&&1.248 & 0.000&  54.62& $-$ \\
			T$_\text{3}$ && 2.022  & 0.000& 1.54& $-$&& 2.515& 0.000& 77.39& $-$&&2.507 & 0.000&  50.52& $-$ \\
			Q$_\text{1}$ && 2.358  & 0.000& 24.78& $-$&& 2.904& 0.000& 68.04& $-$&&2.934 & 0.000&  44.69& $-$ \\
			\bottomrule
			\label{tab:excited_states8910}
		\end{tabular}\\
		\small 
		$^a$ Calculated at SA15-XMCQDPT(8,8)/cc-pVDZ level of theory.\\ $^b$ Calculated at SA15-CASSCF(8,8)/cc-pVDZ level of theory. \\ $^c$ Character of the excited states: GS denotes the ground state. LE$_1$ and LE$_2$ are optically allowed states arising from symmetric and antisymmetric combinations of local excitations within each aza-BODIPY monomer, corresponding to  HOMO→LUMO and  HOMO-1→LUMO transitions, respectively. ME denotes the multiexcitonic state, and CT$_1$ represents charge-transfer states involving electron transfer from the HOMO of one monomer to the LUMO of the other.
	\end{table}
\end{landscape}
Among all the systems studied, the first excited singlet state (S$_{1}$) is an optically bright state with dominant LE$_{1}$ character. In addition to the  LE$_{1}$ states, multiexcitonic (ME) configurations are also present within the low-energy manifold, while the CT$_{1}$ states remain energetically higher than both LE$_{1}$ and ME states. Compared to our previous studies,\cite{Goyal2026} the characterization of the adiabatic states is significantly cleaner, with no noticeable mixing between LE and ME configurations in any of the dimers. Consequently, the ME state remains well separated from the LE manifold, leading to a more distinct SF-relevant state. Furthermore, for all dimers investigated, the lowest quintet state lies above the ME state by approximately 0.1–0.4 eV, providing a more favorable energetic landscape for singlet fission.

\subsection{Diabatic states and couplings}
To evaluate the influence of different spacers on the diabatic electronic couplings, we computed eight adiabatic states for each dimer. These adiabatic states were subsequently transformed into eight corresponding diabatic states using the Truhlar–Nakamura four-fold diabatization scheme.\cite{Nakamura2001, Nakamura2002} The resulting diabatic manifold consists of the following eight electronic states: $^{1}(\textrm{S}_{0}\textrm{S}_{0})$, $^{1}(\textrm{S}_{1}\textrm{S}_{0})$, $^{1}(\textrm{S}_{0}\textrm{S}_{1})$, $^{1}(\textrm{T}_{1}\textrm{T}_{1})$, $^{1}(\textrm{C}\textrm{A})$, $^{1}(\textrm{A}\textrm{C})$, $^{1}(\textrm{DE})_{1}$ and $^{1}(\textrm{DE})_{2}$.

Analysis of the vertical excitation energies (VEEs) obtained from the SA15-XMCQDPT(8,8) and SA8-XMCQDPT(4,4) calculations (see Tables \ref{tab:ortho}–\ref{tab:adamantane}) shows that the (4,4) active space provides a suitable balance between computational efficiency and reliability of the excited-state description. It should be noted that, due to the reduced active space, LE$_\textrm{2}$ type excitations involving the HOMO–1 and LUMO+1 orbitals of the monomers are not included in the diabatization procedure. The resulting diabatic Hamiltonians for all systems are presented in Tables~\ref{tab:diab-ortho}–\ref{tab:diab-adm}, while the corresponding diabatic-to-adiabatic transformation matrices for the nine systems are provided in Eqs.~\ref{eq:rot_ortho}–\ref{eq:rot_adam}. The transformation matrices demonstrate that the adiabatic states possess well-defined electronic characters. In all cases, a distinct ME state with a dominant coefficient could be identified, highlighting the reliability of the diabatization scheme. The adiabatic states exhibiting dominant ME character correspond to states S$_{2}$, S$_{3}$, S$_{2}$, S$_{2}$, S$_{3}$, S$_{2}$, S$_{3}$, S$_{3}$, and S$_{3}$ in Eqs.~\ref{eq:rot_ortho}–\ref{eq:rot_adam}, respectively.

\begin{equation}

	\label{eq:rot_adam}
\end{equation}

\begin{table}[!ht]
	\centering
	\caption{The diabatic electronic energies and couplings (in eVs) between the relevant diabatic states for the system [1] calculated at SA8-XMCQDPT(4,4)/cc-pVDZ level of theory.}
	\begin{tabular}{ccccccccc}
		\toprule
		$\mathcal{H}_{el}$ & $\ket{^{1}(\textrm{S}_{0}\textrm{S}_{0})}$ & $\ket{^{1}(\textrm{S}_{1}\textrm{S}_{0})}$ & $\ket{^{1}(\textrm{S}_{0}\textrm{S}_{1})}$ & $\ket{^{1}(\textrm{T}_{1}\textrm{T}_{1})}$ & $\ket{^{1}(\textrm{C}\textrm{A})}$ & $\ket{^{1}(\textrm{A}\textrm{C})}$ & $\ket{^{1}(\textrm{DE})_{1}}$ & $\ket{^{1}(\textrm{DE})_{2}}$\\
		\midrule		
		$\bra{^{1}(\textrm{S}_{0}\textrm{S}_{0})}$ & 0.0000 & 0.0715 & -0.0733 & -0.0324 & -0.1544 & -0.1558 & 0.5657 & 0.5598 \\
		$\bra{^{1}(\textrm{S}_{1}\textrm{S}_{0})}$ & 0.0715 & 2.0130 & -0.0853 & -0.0043 & 0.2375 & -0.2702 &  0.0103 & -0.0278 \\
		$\bra{^{1}(\textrm{S}_{0}\textrm{S}_{1})}$ & -0.0733 & -0.0853 & 2.0196 & 0.0043 &  0.2649 & -0.2292 & 0.0180 & -0.0147 \\
		$\bra{^{1}(\textrm{T}_{1}\textrm{T}_{1})}$ & -0.0324 & -0.0043 & 0.0043 & 2.4315 & 0.3322 & 0.3321 & 0.1118 & 0.1049 \\
		$\bra{^{1}(\textrm{C}\textrm{A})}$ & -0.1544 & 0.2375 & 0.2649 & 0.3322 & 2.6607 &-0.0272 &  -0.0343 & -0.0272 \\
		$\bra{^{1}(\textrm{A}\textrm{C})}$ & -0.1558 & -0.2702 & -0.2292 & 0.3321 & -0.0272 & 2.6607 & -0.0325 & -0.0325 \\
		$\bra{^{1}(\textrm{DE})_{1}}$ & 0.5657 & 0.0103 & 0.0180 & 0.1118 & -0.0343 & -0.0325 &  3.5576 & -0.1267 \\
		$\bra{^{1}(\textrm{DE})_{2}}$ & 0.5598 & -0.0278 & -0.0147 & 0.1049 & -0.0272 & -0.0325 &  -0.1267 & 3.5609\\
		\bottomrule
	\end{tabular}
	\label{tab:diab-ortho}
\end{table}

\begin{table}[!ht]
	\centering
	\caption{The diabatic electronic energies and couplings (in eVs) between the relevant diabatic states for the system [2] calculated at SA8-XMCQDPT(4,4)/cc-pVDZ level of theory.}
	\begin{tabular}{ccccccccc}
		\toprule
		$\mathcal{H}_{el}$ & $\ket{^{1}(\textrm{S}_{0}\textrm{S}_{0})}$ & $\ket{^{1}(\textrm{S}_{1}\textrm{S}_{0})}$ & $\ket{^{1}(\textrm{S}_{0}\textrm{S}_{1})}$ & $\ket{^{1}(\textrm{T}_{1}\textrm{T}_{1})}$ & $\ket{^{1}(\textrm{C}\textrm{A})}$ & $\ket{^{1}(\textrm{A}\textrm{C})}$ & $\ket{^{1}(\textrm{DE})_{1}}$ & $\ket{^{1}(\textrm{DE})_{2}}$\\
		\midrule
		$\bra{^{1}(\textrm{S}_{0}\textrm{S}_{0})}$ & 0.0000 & 0.0611 & -0.0619 & -0.0057 & -0.0265 & 0.0271 & 0.5396 & 0.5375\\
		$\bra{^{1}(\textrm{S}_{1}\textrm{S}_{0})}$  & 0.0611 & 2.0230 & -0.1340 & 0.0008 & 0.0799 & 0.0061 & -0.0169 & 0.0127\\
		$\bra{^{1}(\textrm{S}_{0}\textrm{S}_{1})}$ & -0.0619 & -0.1340 & 2.0296 & -0.0008 & 0.0032 & 0.0793 & -0.0138 &0.0111\\
		$\bra{^{1}(\textrm{T}_{1}\textrm{T}_{1})}$ & -0.0057 & 0.0008 & -0.0008 & 2.3252 & -0.0678 & 0.0678 & 0.0165 & 0.0153\\
		$\bra{^{1}(\textrm{C}\textrm{A})}$ & 0.0265 & 0.0799 & 0.0032 & -0.0678 & 2.9597 & -0.0431 & 0.0090 & -0.0415\\
		$\bra{^{1}(\textrm{A}\textrm{C})}$ & -0.0271 & 0.0061 & 0.0793 & 0.0678 & -0.0431 & 2.9597 & 0.0408 & -0.0115\\
		$\bra{^{1}(\textrm{DE})_{1}}$ & 0.5396 & -0.0169 & -0.0138 & 0.0165 & 0.0090 & 0.0408 & 3.6506 & -0.1292\\
		$\bra{^{1}(\textrm{DE})_{2}}$ & 0.5375 & 0.0127 & 0.0111 & 0.0153 & -0.0415 & -0.0115 & -0.1292 & 3.6573\\
		\bottomrule	
	\end{tabular}
	\label{tab:diab-meta}
\end{table}

\begin{table}[!ht]
	\centering
	\caption{The diabatic electronic energies and couplings (in eVs) between the relevant diabatic states for the system [3] calculated at SA8-XMCQDPT(4,4)/cc-pVDZ level of theory.}
	\begin{tabular}{ccccccccc}
		\toprule
		$\mathcal{H}_{el}$ & $\ket{^{1}(\textrm{S}_{0}\textrm{S}_{0})}$ & $\ket{^{1}(\textrm{S}_{1}\textrm{S}_{0})}$ & $\ket{^{1}(\textrm{S}_{0}\textrm{S}_{1})}$ & $\ket{^{1}(\textrm{T}_{1}\textrm{T}_{1})}$ & $\ket{^{1}(\textrm{C}\textrm{A})}$ & $\ket{^{1}(\textrm{A}\textrm{C})}$ & $\ket{^{1}(\textrm{DE})_{1}}$ & $\ket{^{1}(\textrm{DE})_{2}}$\\
		\midrule	
		$\bra{^{1}(\textrm{S}_{0}\textrm{S}_{0})}$ & 0.0000 & 0.0537& -0.0549 & 0.0537 & -0.1847 & 0.1859 & 0.5658 & 0.5598 \\
		$\bra{^{1}(\textrm{S}_{1}\textrm{S}_{0})}$ & 0.0537 & 2.0495 & 0.1402 & -0.0025 & 0.3040 & 0.2888 & -0.0126 & -0.0171 \\
		$\bra{^{1}(\textrm{S}_{0}\textrm{S}_{1})}$ &-0.0549 & -0.1402 & 2.0562 & 0.0025 & 0.2829 & 0.2957 & 0.0109 & 0.0096\\
		$\bra{^{1}(\textrm{T}_{1}\textrm{T}_{1})}$ & 0.0537 & -0.0025 & 0.0025 & 2.2953 & -0.3934 & 0.3935 & -0.1499 & -0.1403\\
		$\bra{^{1}(\textrm{C}\textrm{A})}$ & -0.1847 & 0.3040 & 0.2829 & -0.3934 & 2.8601 & 0.0516 &  -0.0442 & -0.0381\\
		$\bra{^{1}(\textrm{A}\textrm{C})}$ &0.1859 & 0.2888 & 0.2957 & 0.3935 & 0.0516 & 2.8601 &  0.0445 & 0.0410\\
		$\bra{^{1}(\textrm{DE})_{1}}$ & 0.5658 & -0.0126 & 0.0109 & -0.1499 & -0.0442 & 0.0445 &  3.4546 & -0.1570\\
		$\bra{^{1}(\textrm{DE})_{2}}$ & 0.5598 & -0.0171 & 0.0096 & -0.1403 & -0.0381 & 0.0410 &  -0.1570 & 3.4646\\
		\bottomrule	
	\end{tabular}
	\label{tab:diab-para}
\end{table}

\begin{table}[!ht]
	\centering
	\caption[The diabatic electronic energies and couplings (in eVs) between the diabatic states for the system {[}5{]}.]{The diabatic electronic energies and couplings (in eVs) between the relevant diabatic states for the system [4] calculated at SA8-XMCQDPT(4,4)/cc-pVDZ level of theory.}
	\begin{tabular}{ccccccccc}
		\toprule
		$\mathcal{H}_{el}$ & $\ket{^{1}(\textrm{S}_{0}\textrm{S}_{0})}$ & $\ket{^{1}(\textrm{S}_{1}\textrm{S}_{0})}$ & $\ket{^{1}(\textrm{S}_{0}\textrm{S}_{1})}$ & $\ket{^{1}(\textrm{T}_{1}\textrm{T}_{1})}$ & $\ket{^{1}(\textrm{C}\textrm{A})}$ & $\ket{^{1}(\textrm{A}\textrm{C})}$ & $\ket{^{1}(\textrm{DE})_{1}}$ & $\ket{^{1}(\textrm{DE})_{2}}$\\
		\midrule
		$\bra{^{1}(\textrm{S}_{0}\textrm{S}_{0})}$ & 0.0000 & -0.0079 &  0.0083 & -0.0161 & -0.0983 & -0.0984 &  0.5697 &  0.5646 \\
		$\bra{^{1}(\textrm{S}_{1}\textrm{S}_{0})}$ &-0.0079 &  1.9864 & -0.0974 &  0.0008 &  0.1582 & -0.1463 & -0.0277 &  0.0000 \\
		$\bra{^{1}(\textrm{S}_{0}\textrm{S}_{1})}$ &0.0083 & -0.0974 &  1.9897 & -0.0008 &  0.1435 & -0.1554 &  0.0010 &  0.0276 \\
		$\bra{^{1}(\textrm{T}_{1}\textrm{T}_{1})}$ &-0.0161 &  0.0008 & -0.0008 &  2.1226 &  0.2235 &  0.2235 &  0.0398 &  0.0368 \\
		$\bra{^{1}(\textrm{C}\textrm{A})}$ &-0.0983 &  0.1582 &  0.1435 &  0.2235 &  3.0660 & -0.0193 & -0.0157 & -0.0199 \\
		$\bra{^{1}(\textrm{A}\textrm{C})}$ &-0.0984 & -0.1463 & -0.1554 &  0.2235 & -0.0193 &  3.0660 & -0.0223 & -0.0131 \\
		$\bra{^{1}(\textrm{DE})_{1}}$ &0.5697 & -0.0277 &  0.0010 &  0.0398 & -0.0157 & -0.0223 &  3.4115 & -0.1487 \\
		$\bra{^{1}(\textrm{DE})_{2}}$ &0.5646 &  0.0000 &  0.0276 &  0.0368 & -0.0199 & -0.0131 & -0.1487 &  3.4214 \\
		\bottomrule	
	\end{tabular}
	\label{tab:diab-acet-ortho}
\end{table}

\begin{table}[!ht]
	\centering
	\caption{The diabatic electronic energies and couplings (in eVs) between the relevant diabatic states for the system [5] calculated at SA8-XMCQDPT(4,4)/cc-pVDZ level of theory.}
	\begin{tabular}{ccccccccc}
		\toprule
		$\mathcal{H}_{el}$ & $\ket{^{1}(\textrm{S}_{0}\textrm{S}_{0})}$ & $\ket{^{1}(\textrm{S}_{1}\textrm{S}_{0})}$ & $\ket{^{1}(\textrm{S}_{0}\textrm{S}_{1})}$ & $\ket{^{1}(\textrm{T}_{1}\textrm{T}_{1})}$ & $\ket{^{1}(\textrm{C}\textrm{A})}$ & $\ket{^{1}(\textrm{A}\textrm{C})}$ & $\ket{^{1}(\textrm{DE})_{1}}$ & $\ket{^{1}(\textrm{DE})_{2}}$\\
		\midrule
		$\ket{^{1}(\textrm{S}_{0}\textrm{S}_{0})}$ &	0.0000     & -0.0052 &  0.0051 & -0.0026 &  0.5668 &  0.0290  & -0.0288 &  0.5633 \\
		$\bra{^{1}(\textrm{S}_{1}\textrm{S}_{0})}$ & -0.0052 &  1.9897 & -0.0802 & -0.0001 &  0.0168 & -0.0588 & -0.0087 &  0.003  \\
		$\bra{^{1}(\textrm{S}_{0}\textrm{S}_{1})}$ &  0.0051 & -0.0802 &  1.9897 &  0.0002 & -0.0022 & -0.0102 & -0.0585 & -0.0168 \\
		$\bra{^{1}(\textrm{T}_{1}\textrm{T}_{1})}$ &	 -0.0026 & -0.0001 &  0.0002 &  2.1359 &  0.0051 & -0.0418 &  0.0416 &  0.0047 \\
		$\bra{^{1}(\textrm{C}\textrm{A})}$ &  	0.5668 &  0.0168 & -0.0022 &  0.0051 &  3.4879 & -0.0186 &  0.0020  & -0.1435 \\
		$\bra{^{1}(\textrm{A}\textrm{C})}$ &	0.0290  & -0.0588 & -0.0102 & -0.0418 & -0.0186 &  3.4912 & -0.0116 & -0.0008 \\
		$\bra{^{1}(\textrm{DE})_{1}}$& -0.0288 & -0.0087 & -0.0585 &  0.0416 &  0.0020  & -0.0116 &  3.4912 &  0.0188 \\
		$\bra{^{1}(\textrm{DE})_{2}}$& 0.5633 &  0.0030  & -0.0168 &  0.0047 & -0.1435 & -0.0008 &  0.0188 &  3.4978 \\
		\bottomrule	
	\end{tabular}
	\label{tab:diab-acet-meta}
\end{table}

\begin{table}[!ht]
	\centering
	\caption{The diabatic electronic energies and couplings (in eVs) between the relevant diabatic states for the system [6] calculated at SA8-XMCQDPT(4,4)/cc-pVDZ level of theory.}
	\begin{tabular}{ccccccccc}
		\toprule
		$\mathcal{H}_{el}$ & $\ket{^{1}(\textrm{S}_{0}\textrm{S}_{0})}$ & $\ket{^{1}(\textrm{S}_{1}\textrm{S}_{0})}$ & $\ket{^{1}(\textrm{S}_{0}\textrm{S}_{1})}$ & $\ket{^{1}(\textrm{T}_{1}\textrm{T}_{1})}$ & $\ket{^{1}(\textrm{C}\textrm{A})}$ & $\ket{^{1}(\textrm{A}\textrm{C})}$ & $\ket{^{1}(\textrm{DE})_{1}}$ & $\ket{^{1}(\textrm{DE})_{2}}$\\
		\midrule
		$\bra{^{1}(\textrm{S}_{0}\textrm{S}_{0})}$ & 0.0000 & -0.0121 &  0.0117 & -0.0286 & -0.1366 &  0.1369 &  0.5707 &  0.5656 \\
		$\bra{^{1}(\textrm{S}_{1}\textrm{S}_{0})}$& -0.0121 &  1.9931 & -0.0913 &  0.0021 &  0.2340 &  0.1992 & -0.0009 & -0.0310 \\
		$\bra{^{1}(\textrm{S}_{0}\textrm{S}_{1})}$ & 0.0117 & -0.0913 &  1.9931 & -0.0018 &  0.1965 &  0.2303 &  0.0309 & -0.0025 \\
		$\bra{^{1}(\textrm{T}_{1}\textrm{T}_{1})}$ & -0.0286 &  0.0021 & -0.0018 &  2.1193 &  0.3049 & -0.3047 &  0.0697 &  0.0642 \\
		$\bra{^{1}(\textrm{C}\textrm{A})}$ & -0.1366 &  0.2340 &  0.1965 &  0.3049 &  3.3085 &  0.0391 & -0.0200 & -0.0163 \\
		$\bra{^{1}(\textrm{A}\textrm{C})}$ & 0.1369 &  0.1992 &  0.2303 & -0.3047 &  0.0391 &  3.3085 &  0.0195 &  0.0172 \\
		$\bra{^{1}(\textrm{DE})_{1}}$ & 0.5707 & -0.0009 &  0.0309 &  0.0697 & -0.0200 &  0.0195 &  3.3915 & -0.1464 \\
		$\bra{^{1}(\textrm{DE})_{2}}$ & 0.5656 & -0.0310 & -0.0025 &  0.0642 & -0.0163 &  0.0172 & -0.1464 &  3.4015 \\
		\bottomrule
	\end{tabular}
	\label{tab:diab-acet-para}
\end{table}

\begin{table}[!ht]
	\centering
	\caption[The diabatic electronic energies and couplings (in eVs) between the diabatic states for the system {[}8{]}.]{The diabatic electronic energies and couplings (in eVs) between the relevant diabatic states for the system [7] calculated at SA8-XMCQDPT(4,4)/cc-pVDZ level of theory.}
	\begin{tabular}{ccccccccc}
		\toprule
		$\mathcal{H}_{el}$ & $\ket{^{1}(\textrm{S}_{0}\textrm{S}_{0})}$ & $\ket{^{1}(\textrm{S}_{1}\textrm{S}_{0})}$ & $\ket{^{1}(\textrm{S}_{0}\textrm{S}_{1})}$ & $\ket{^{1}(\textrm{T}_{1}\textrm{T}_{1})}$ & $\ket{^{1}(\textrm{C}\textrm{A})}$ & $\ket{^{1}(\textrm{A}\textrm{C})}$ & $\ket{^{1}(\textrm{DE})_{1}}$ & $\ket{^{1}(\textrm{DE})_{2}}$\\
		\midrule
		$\bra{^{1}(\textrm{S}_{0}\textrm{S}_{0})}$ & 0.0000 & 0.0406 & -0.0409 & 0.0056 & -0.0676 & 0.0681 &  -0.5286 & -0.5251 \\
		$\bra{^{1}(\textrm{S}_{1}\textrm{S}_{0})}$ & 0.0406 & 1.8735 & -0.0754 & -0.0023 & 0.2010 & -0.0218 &  0.0143 & 0.013 \\
		$\bra{^{1}(\textrm{S}_{0}\textrm{S}_{1})}$ & -0.0409 & -0.0754 & 1.8801 & 0.0031 & -0.0303 & 0.2023 & -0.0079 & -0.0137 \\
		$\bra{^{1}(\textrm{T}_{1}\textrm{T}_{1})}$  & 0.0056 & -0.0023 & 0.0031 & 2.0529 & -0.1290 & 0.1296 &  -0.0212 & -0.0188 \\
		$\bra{^{1}(\textrm{C}\textrm{A})}$ & -0.0676 & 0.2010 & -0.0303 & -0.1290 & 2.8434 & -0.0339 &  0.0571 & -0.0411 \\
		$\bra{^{1}(\textrm{A}\textrm{C})}$ & 0.0681 & -0.0218 & 0.2023 & 0.1296 & -0.0339 & 2.8434 & 0.0357 & -0.0592 \\
		$\bra{^{1}(\textrm{DE})_{1}}$ & -0.5286 & 0.0143 & -0.0079 & -0.0212 & 0.0571 & 0.0357 &  3.2553 & -0.1496 \\
		$\bra{^{1}(\textrm{DE})_{2}}$ & -0.5251 & 0.0130 & -0.0137 & -0.0188 & -0.0411 & -0.0592 &  -0.1496 & 3.2686 \\
		\bottomrule
	\end{tabular}
	\label{tab:diab-xc}
\end{table}

\begin{table}[!ht]
	\centering
	\caption[The diabatic electronic energies and couplings (in eVs) between the diabatic states for the system {[}9{]}.]{The diabatic electronic energies and couplings (in eVs) between the relevant diabatic states for the system [8] calculated at SA8-XMCQDPT(4,4)/cc-pVDZ level of theory.}
	\begin{tabular}{ccccccccc}
		\toprule
		$\mathcal{H}_{el}$ & $\ket{^{1}(\textrm{S}_{0}\textrm{S}_{0})}$ & $\ket{^{1}(\textrm{S}_{1}\textrm{S}_{0})}$ & $\ket{^{1}(\textrm{S}_{0}\textrm{S}_{1})}$ & $\ket{^{1}(\textrm{T}_{1}\textrm{T}_{1})}$ & $\ket{^{1}(\textrm{C}\textrm{A})}$ & $\ket{^{1}(\textrm{A}\textrm{C})}$ & $\ket{^{1}(\textrm{DE})_{1}}$ & $\ket{^{1}(\textrm{DE})_{2}}$\\
		\midrule
		$\bra{^{1}(\textrm{S}_{0}\textrm{S}_{0})}$ & 0.0000 & 0.0691 & 0.0708 & 0.0180 & 0.1188 & -0.1203 & -0.4815 & -0.4778 \\
		$\bra{^{1}(\textrm{S}_{1}\textrm{S}_{0})}$	& 0.0691 & 1.8205 & 0.0377 & 0.0021 & -0.2588 & -0.2388 & -0.0402 & 0.0226 \\
		$\bra{^{1}(\textrm{S}_{0}\textrm{S}_{1})}$ & 0.0708 & 0.0377 & 1.8236 & 0.0027 & 0.2332 & 0.2503 & 0.0106 & -0.0444 \\
		$\bra{^{1}(\textrm{T}_{1}\textrm{T}_{1})}$ & 0.0180 & 0.0021 & 0.0027 & 2.4306 & 0.3181 & -0.3179 & 0.1075 & 0.1011 \\
		$\bra{^{1}(\textrm{C}\textrm{A})}$ & 0.1188 & -0.2588 & 0.2332 & 0.3181 & 2.5725 & 0.0128 & -0.0458 & -0.0253 \\
		$\bra{^{1}(\textrm{A}\textrm{C})}$ & -0.1203 & -0.2388 & 0.2503 & -0.3179 & 0.0128 & 2.5731 & 0.0311 & 0.0446 \\
		$\bra{^{1}(\textrm{DE})_{1}}$ & -0.4815 & -0.0402 & 0.0106 & 0.1075 & -0.0458 & 0.0311 & 3.2282 & -0.0865 \\
		$\bra{^{1}(\textrm{DE})_{2}}$ & -0.4778 & 0.0226 & -0.0444 & 0.1011 & -0.0253 & 0.0446 & -0.0865 & 3.2316 \\
		\bottomrule
	\end{tabular}
	\label{tab:diab-xc-no-acet}
\end{table}

\begin{table}[!ht]
	\centering
	\caption[The diabatic electronic energies and couplings (in eVs) between the diabatic states for the system {[}10{]}.]{The diabatic electronic energies and couplings (in eVs) between the relevant diabatic states for the system [9] calculated at SA8-XMCQDPT(4,4)/cc-pVDZ level of theory.}
	\begin{tabular}{ccccccccc}
		\toprule
		$\mathcal{H}_{el}$ & $\ket{^{1}(\textrm{S}_{0}\textrm{S}_{0})}$ & $\ket{^{1}(\textrm{S}_{1}\textrm{S}_{0})}$ & $\ket{^{1}(\textrm{S}_{0}\textrm{S}_{1})}$ & $\ket{^{1}(\textrm{T}_{1}\textrm{T}_{1})}$ & $\ket{^{1}(\textrm{DE})_{1}}$ & $\ket{^{1}(\textrm{DE})_{2}}$& $\ket{^{1}(\textrm{C}\textrm{A})}$ & $\ket{^{1}(\textrm{A}\textrm{C})}$ \\
		\midrule
		$\bra{^{1}(\textrm{S}_{0}\textrm{S}_{0})}$ & 0.0000 & -0.0389 & -0.0374 & 0.0000 & -0.5687 & 0.5675 & -0.0013 & -0.0014 \\
		$\bra{^{1}(\textrm{S}_{1}\textrm{S}_{0})}$ & -0.0389 & 2.0097 & 0.0412 & 0.0000 & -0.0070 & 0.0119 & 0.0021 & -0.0015 \\
		$\bra{^{1}(\textrm{S}_{0}\textrm{S}_{1})}$ & -0.0374 & 0.0412 & 2.0130 & 0.0000 & -0.0051 & 0.0073 & -0.0017 & 0.0021 \\
		$\bra{^{1}(\textrm{T}_{1}\textrm{T}_{1})}$  & 0.0000 & 0.0000 & 0.0000 & 2.2057 & 0.0000 & 0.0000 & 0.0022 & 0.0022 \\
		$\bra{^{1}(\textrm{DE})_{1}}$ & -0.5687 & -0.0070 & -0.0051 & 0.0000 & 3.6075 & 0.1462 & -0.0002 & -0.0002 \\
		$\bra{^{1}(\textrm{DE})_{2}}$ & 0.5675 & 0.0119 & 0.0073 & 0.0000 & 0.1462 & 3.6207 & 0.0002 & 0.0002 \\
		$\bra{^{1}(\textrm{C}\textrm{A})}$ & -0.0013 & 0.0021 & -0.0017 & 0.0022 & -0.0002 & 0.0002 & 3.8832 & 0.0000 \\
		$\bra{^{1}(\textrm{A}\textrm{C})}$ & -0.0014 & -0.0015 & 0.0021 & 0.0022 & -0.0002 & 0.0002 & 0.0000 & 3.8898 \\
		\bottomrule
	\end{tabular}
	\label{tab:diab-adm}
\end{table}

\begin{table}[!ht]
	\centering
	\caption[Diabatic electronic couplings among the states $^1\left(\textrm{S}_\textrm{1} \textrm{S}_\textrm{0}\right)$,  $^1\left(\textrm{S}_\textrm{0} \textrm{S}_\textrm{1}\right)$,  $^1\left(\textrm{T}_\textrm{1} \textrm{T}_\textrm{1}\right)$, $^{1}(\textrm{C}\textrm{A})$ and $^{1}(\textrm{A}\textrm{C})$ for all the dimer systems {[}1{]} to {[}10{]}.]{Diabatic electronic couplings among the states $^1\left(\textrm{S}_\textrm{1} \textrm{S}_\textrm{0}\right)$,  $^1\left(\textrm{S}_\textrm{0} \textrm{S}_\textrm{1}\right)$,  $^1\left(\textrm{T}_\textrm{1} \textrm{T}_\textrm{1}\right)$, $^{1}(\textrm{C}\textrm{A})$ and $^{1}(\textrm{A}\textrm{C})$ for all the dimer systems [1] to [9], expressed in meV, as obtained via Truhlar’s fourfold diabatization approach.}
	\small
	\begin{center}
		\begin{tabular}{rrrrrrrrrr}
			\toprule
			& [1] &  [2] & [3] &[4] & [5] & [6] & [7] & [8] & [9] \\
			\toprule
			$\bra{^{1}(\textrm{S}_\textrm{1}\textrm{S}_\textrm{0})} V \ket{^{1}(\textrm{S}_\textrm{0}\textrm{S}_\textrm{1})}$    & -85& -134&-140& -97& -80 &-91& -75& 38& 41\\
			$\bra{^{1}(\textrm{S}_\textrm{1}\textrm{S}_\textrm{0})} V \ket{^{1}(\textrm{T}_\textrm{1}\textrm{T}_\textrm{1})}$    & -4&  1& -2& 1& 0& 2&  -2& 2& 0\\
			$\bra{^{1}(\textrm{S}_\textrm{1}\textrm{S}_\textrm{0})} V \ket{^{1}(\textrm{C}\textrm{A})}$            & 237& 80 & 304& 158& 17& 234&  201& -259& 2\\
			$\bra{^{1}(\textrm{S}_\textrm{1}\textrm{S}_\textrm{0})} V \ket{^{1}(\textrm{A}\textrm{C})}$            & -270& 6 & 289& -146& -59& 199&  -22& -239& -1\\
			$\bra{^{1}(\textrm{S}_\textrm{0}\textrm{S}_\textrm{1})} V \ket{^{1}(\textrm{T}_\textrm{1}\textrm{T}_\textrm{1})}$   & 4 & -1& 2&  -1& 0 & -2&  3& 3& 0\\
			$\bra{^{1}(\textrm{S}_\textrm{0}\textrm{S}_\textrm{1})} V \ket{^{1}(\textrm{C}\textrm{A})}$            & 265& 3& 283& 143& -2& 196&  -30& 233& -2\\
			$\bra{^{1}(\textrm{S}_\textrm{0}\textrm{S}_\textrm{1})} V \ket{^{1}(\textrm{A}\textrm{C})}$            & -229& 79 & 296& -155& -10& 230&  202& 250& 2\\
			$\bra{^{1}(\textrm{T}_\textrm{1}\textrm{T}_\textrm{1})} V \ket{^{1}(\textrm{C}\textrm{A})}$            & 332& -68 & -393& 223& 5& 305&  -130& 318& 2 \\
			$\bra{^{1}(\textrm{T}_\textrm{1}\textrm{T}_\textrm{1})} V \ket{^{1}(\textrm{A}\textrm{C})}$            & 332& 68 & 393& 223& -42& -305&  130& -318& 2\\
			\bottomrule
		\end{tabular}
	\end{center}
	\label{tab:eff-spacer}
\end{table}

\vspace{10cm}
The key diabatic couplings for all dimers are summarized in Table \ref{tab:eff-spacer}. The direct coupling $\bra{^{1}(\textrm{S}_\textrm{1}\textrm{S}_\textrm{0})} V \ket{^{1}(\textrm{T}_\textrm{1}\textrm{T}_\textrm{1})}$ is negligible for every system, indicating weak direct interaction between the LE and ME diabatic states in the spacer-linked aza-BODIPY dimers similar to the tetracene dimers.\cite{Sebastian1981} In contrast, the mediated couplings ($\bra{^{1}(\textrm{S}_\textrm{1}\textrm{S}_\textrm{0})} V \ket{^{1}(\textrm{C}\textrm{A})}$ and  $\bra{^{1}(\textrm{T}_\textrm{1}\textrm{T}_\textrm{1})} V \ket{^{1}(\textrm{C}\textrm{A})}$) are generally strong for most systems, suggesting that population transfer from LE to ME states proceeds predominantly through an indirect CT-mediated pathway. Systems [7] and [9] differ from the remaining dimers in exhibiting generally weak diabatic couplings between all states. In particular, the non-conjugated character of system [9] nearly suppresses electronic communication entirely, resulting in minimal interstate population transfer.

\begin{table}[!ht]
	\small
	\centering
	\caption[Diabatic direct, mediated, and effective couplings (in meV) for the investigated dimers.]{Diabatic direct, mediated, and effective couplings (in meV) for the investigated dimers. The SF rate constant is estimated assuming a reorganization energy of 100 meV for the $\ket{^{1}(\textrm{S}_\textrm{1}\textrm{S}_\textrm{0})}$ state.}
	\begin{center}
		\begin{tabular}{crrrc}
			\toprule
			System & Direct & Mediated & V$_\textrm{eff}$ & k$_\textrm{SF}^a$ \\
			& (meV)  & (meV)    &   (meV)          & (s$^\textrm{-1}$)\\
			\toprule
			1  &  -4.4& 24.9& 20.5 &   8.17 $\times$ 10$^\textrm{9}$  \\
			2  & 0.8 & 6.4& 5.6 &  1.77$\times$ 10$^\textrm{4}$   \\
			3  & -2.5 & 8.7& 6.2 &   6.51$\times$ 10$^\textrm{4}$  \\
			4 & 0.8 & 2.6 & -3.4 &  1.26$\times$ 10$^\textrm{-23}$  \\
			5  & 0.1 & -1.8& -1.9&  3.02$\times$ 10$^\textrm{-61}$   \\
			6  & 1.9 & -105.4& -107.5 &  9.57 $\times$ 10$^\textrm{-43}$  \\
			7  & -2.3 & 32.7& 30.4 &  1.80 $\times$ 10$^\textrm{-8}$   \\
			8 & 2.1 & 14.3& 16.4 &    0.48 \\
			9  & 0.0 & 18.6& 18.6 &   1.06 $\times$ 10$^\textrm{-92}$  \\
			\bottomrule
		\end{tabular} \\
		$^a$ $E(\ket{^{1}(\textrm{T}_\textrm{1}\textrm{T}_\textrm{1})})-E(\ket{^{1}(\textrm{S}_\textrm{1}\textrm{S}_\textrm{0})})$ taken from Tables \ref{tab:diab-ortho} - \ref{tab:diab-adm}.\\ 
	\end{center}
	\label{tab:rate-const-spacer}
\end{table}
\subsection{Factors Governing Singlet-Fission Activity in Aza-BODIPY Dimers }
We also computed the SF rate constants, $k_\textrm{SF}$ for all dimers (Table \ref{tab:rate-const-spacer}), and three clear structure-property trends emerge:
\begin{itemize}
	\item[3.3.1] \textbf{Effect of substitution pattern on singlet fission activity :}\\
	To understand the influence of the substitution pattern on SF activity, we first analyzed the ortho-, meta-, and para-linked benzene spacer dimers without acetylene units (systems 1,2 and 3). The calculated SF rates follow the trend
	$k_\textrm{SF}$(ortho) >> 	$k_\textrm{SF}$(para) > $k_\textrm{SF}$(meta)
	Interestingly, this behavior cannot be explained solely by the LE–ME energy gap, since the para-linked system exhibits the smallest LE–ME separation among the three dimers. Instead, the dominant factor governing the SF activity is the CT-mediated indirect coupling pathway. In all three dimers, the direct LE–ME coupling is negligible, indicating that population transfer from the LE state to the ME state occurs primarily through intermediate CT states.
	
	The ortho-linked dimer exhibits a significantly smaller ME–CT energy gap compared to the para- and meta-linked systems. This energetic alignment strongly enhances the indirect coupling pathway and leads to a much larger effective coupling, resulting in an exceptionally high SF rate. In contrast, the larger ME–CT separations in the para- and meta-linked dimers suppress the CT-mediated interaction and consequently reduce $k_\textrm{SF}$.
	
	The substitution pattern further influences the efficiency of electronic communication through the benzene spacer. While ortho- and para-linkages preserve effective $\pi$-conjugation between the aza-BODIPY units, the meta connectivity disrupts the conjugation pathway, leading to weaker electronic communication and smaller effective couplings. In addition, the moderate intermonomer distance in the ortho-linked system enhances orbital overlap between the chromophores, which further increases the effective coupling. Consequently, the ortho-linked dimer exhibits the highest SF activity, whereas the para-linked system shows only moderately larger SF rates than the meta analogue due to its more efficient $\pi$-conjugative interaction. 
	
	\item[3.3.2] \textbf{Elongation of the spacer through acetylene units:}\\
	To further investigate the role of intermonomer separation and conjugation length, acetylene units were introduced between the benzene spacer and aza-BODIPY chromophores (systems 4, 5 and 6) similar to pentacene system.\cite{Zirzlmeier2015} For these acetylene-linked dimers, the calculated SF rates decrease dramatically for all substitution patterns compared to the directly linked benzene spacers. Despite this suppression, the relative trend
	$k_\textrm{SF}$(ortho) > $k_\textrm{SF}$(para) > $k_\textrm{SF}$(meta)
	
	is preserved.
	
	A notable observation is that the LE–ME energy gaps become uniformly small in all acetylene-linked dimers. Therefore, the substantial reduction in $k_\textrm{SF}$ cannot be attributed to unfavorable LE–ME energetics. Instead, the dominant factor is again the CT-mediated indirect coupling mechanism. Upon introduction of the acetylene units, the ME–CT energy gap increases significantly for all three substitution patterns, indicating that the CT states become energetically less accessible for mediating the LE-to-ME population transfer. As a consequence, the effective coupling decreases drastically, leading to extremely small SF rates. 
	
	Although acetylene units extend the $\pi$-conjugation length, they simultaneously increase the intermonomer separation between the aza-BODIPY units. The larger separation reduces orbital overlap and weakens the electronic communication between the chromophores, thereby suppressing the CT-mediated interaction responsible for efficient SF. Consequently, the effect of increased intermonomer distance dominates over the beneficial effect of extended conjugation in these systems.
	
	Within the acetylene-linked series, the ortho-linked dimer still exhibits the largest SF rate because of its relatively shorter intermonomer distance and stronger electronic communication compared to the para- and meta-linked analogues. The para-linked system retains more efficient $\pi$-conjugation than the meta-linked dimer and therefore shows slightly larger effective couplings and SF rates.
	
	This trend is further supported by the comparison between the cross-conjugated systems [8] and [7], where system [7] contains acetylene units while system [8] does not. Despite their similar conjugation topology, the calculated $k_\textrm{SF}$ decreases dramatically from 0.48 in system [8] to 1.8×10$^\textrm{-8}$ in system [7]. This further confirms that the introduction of acetylene units suppresses SF activity primarily by increasing the intermonomer separation and weakening the CT-mediated electronic communication between the aza-BODIPY units.

   Overall, these results demonstrate that efficient SF in aza-BODIPY dimers requires not only favorable LE–ME energetics but, more importantly, strong CT-mediated coupling facilitated by optimal intermonomer separation and effective electronic communication between the chromophores.
	
\item[3.3.3] \textbf{Effect of conjugation on singlet fission activity:}\\
To further understand the role of electronic communication between the chromophores, we compared fully conjugated,\cite{Zirzlmeier2015} cross-conjugated,\cite{Zirzlmeier2016} and non-conjugated\cite{Basel2017,Basel2018} aza-BODIPY dimers represented by systems [1], [8], and [9], respectively. The calculated SF rates show a dramatic decrease with the reduction of conjugation,
	 \(k_{\mathrm{SF}}(conjugated) > k_{\mathrm{SF}}(cross-conjugated) > k_{\mathrm{SF}}(non-conjugated)\) with values of 8.17 $\times$ 10$^\textrm{9}$, 0.48 and 1.06 $\times$ 10$^\textrm{-92}$, respectively.
	 
	The fully conjugated ortho-linked benzene spacer system exhibits the highest SF activity due to its strong electronic communication and favorable CT-mediated indirect coupling. In contrast, the cross-conjugated system [8] shows a substantial reduction in $k_\textrm{SF}$, despite possessing a relatively small ME–CT energy gap. This indicates that partial disruption of conjugation weakens the effective electronic interaction between the aza-BODIPY units and reduces the efficiency of the indirect coupling pathway.
	
	The most striking behavior is observed for the non-conjugated adamantane-linked system [9]. Although this dimer possesses a comparatively small LE–ME energy gap, both the direct and indirect couplings become negligibly small because the adamantane spacer completely interrupts $\pi$-conjugation between the chromophores. In addition, the large ME–CT energy separation further suppresses the CT-mediated interaction, leading to an almost vanishing SF rate. These results clearly demonstrate that efficient SF in aza-BODIPY dimers requires not only favorable energetic alignment among the LE, CT, and ME states, but also sufficient electronic communication mediated through conjugated pathways.
	 
\end{itemize}

\section*{Conclusions}

In this work, we systematically investigated the singlet-fission (SF) activity of spacer-linked aza-BODIPY dimers using a diabatic electronic-structure framework. Our results reveal that the SF efficiency in these systems is governed primarily by the relative energetic alignment of the locally excited (LE), charge-transfer (CT), and multiexcitonic (ME) states together with the strength of the CT-mediated indirect coupling pathway. Since the direct LE–ME coupling is negligible in all dimers studied, efficient SF depends critically on the ability of the CT states to mediate the interaction between LE and ME configurations.

Among the different substitution patterns, the ortho-linked dimers exhibit the highest SF activity owing to their smaller ME–CT energy separation, stronger effective couplings, and moderate intermonomer distances, which collectively enhance electronic communication between the chromophores. In contrast, the weaker $\pi$-conjugative interaction in the meta-linked systems leads to reduced effective couplings and lower SF rates, while the para-linked dimers display intermediate behavior.

The introduction of acetylene units strongly suppresses SF activity, despite producing comparatively favorable LE–ME energy gaps. This behavior originates from the increased intermonomer separation and the resulting reduction in CT-mediated electronic communication, which substantially weakens the effective coupling between the SF-relevant states. Furthermore, the comparison between conjugated, cross-conjugated, and non-conjugated spacers demonstrates that efficient electronic communication through the spacer is essential for sustaining strong SF activity. In particular, the non-conjugated adamantane-linked system exhibits negligibly small coupling elements and an almost vanishing SF rate.

Overall, this study establishes clear structure-property relationships connecting substitution pattern, conjugation, intermonomer separation, and diabatic state energetics with SF efficiency in aza-BODIPY dimers. These findings provide important molecular design principles for developing BODIPY-based chromophores with improved triplet-generation capabilities and enhanced SF performance.

\section*{Conflicts of interest}
There are no conflicts to declare.

\section*{Acknowledgements}
S. G. expresses gratitude to the University Grants Commission (UGC), India, for providing the Senior Research Fellowship. SRR acknowledges financial support from SERB, India, for computational resources under the project (SRG/2021/001684), the UGC Startup Grant (F.30-547/2021(BSR)) and ANRF pair grant (ANRF/PAIR/2025/000006/PAIR-A).

\bibliography{ref}

@Article{Roncali2009,
	author    = {Roncali, Jean},
	title     = {Molecular {B}ulk {H}eterojunctions: {A}n {E}merging {A}pproach to {O}rganic {S}olar {C}ells},
	journal   = {Acc Chem Res},
	year      = {2009},
	volume    = {42},
	number    = {11},
	pages     = {1719--1730},
	doi       = {10.1021/ar900041b},
	publisher = {ACS Publications},
}

@Article{Reddy2024,
  author   = {Reddy, S. Rajagopala and Coto, Pedro B. and Thoss, Michael},
  journal  = {J. Chem. Phys.},
  title    = {Intramolecular Singlet Fission: Quantum Dynamical Simulations Including the Effect of the Laser Field},
  year     = {2024},
  issn     = {0021-9606},
  month    = {05},
  number   = {19},
  pages    = {194306},
  volume   = {160},
  doi      = {10.1063/5.0209546},
  eprint   = {https://pubs.aip.org/aip/jcp/article-pdf/doi/10.1063/5.0209546/19959388/194306\_1\_5.0209546.pdf},
}

@Article{Squeo2020,
  author    = {Squeo, Benedetta Maria and Ganzer, Lucia and Virgili, Tersilla and Pasini, Mariacecilia},
  title     = {Bodipy-based Molecules, a Platform for Photonic and Solar Cells},
  journal   = {Molecules},
  year      = {2020},
  volume    = {26},
  number    = {1},
  pages     = {153},
  doi       = {10.3390/molecules26010153},
  publisher = {MDPI},
}

@Article{Madhu2011,
  author    = {Madhu, Sheri and Rao, Malakalapalli Rajeswara and Shaikh, Mushtaque S and Ravikanth, Mangalampalli},
  title     = {3, 5-diformylboron Dipyrromethenes As Fluorescent Ph Sensors},
  journal   = {Inorg. Chem.},
  year      = {2011},
  volume    = {50},
  number    = {10},
  pages     = {4392--4400},
  doi       = {10.1021/ic102499h},
  publisher = {ACS Publications},
}

@Article{Reddy2022,
  author  = {Papadopoulos, Ilias and Reddy, S. Rajagopala and Coto, Pedro B. and Lehnherr, Dan and Thiel, Dominik and Thoss, Michael and Tykwinski, Rik R. and Guldi, Dirk M.},
  journal = {J. Phys. Chem. Lett.},
  title   = {Parallel Versus Twisted Pentacenes: Conformational Impact on Singlet Fission},
  year    = {2022},
  number  = {23},
  pages   = {5094-5100},
  volume  = {13},
  doi     = {10.1021/acs.jpclett.2c01395},
}

@Article{Stephens1994,
  author    = {Stephens, Philip J and Devlin, Frank J and Chabalowski, Cary F and Frisch, Michael J},
  title     = {Ab Initio Calculation of Vibrational Absorption and Circular Dichroism Spectra Using Density Functional Force Fields},
  journal   = {J. Phys. Chem.},
  year      = {1994},
  volume    = {98},
  number    = {45},
  pages     = {11623--11627},
  doi       = {10.1021/j100096a001},
  publisher = {ACS Publications},
}

@Article{Niu2013,
  author    = {Niu, Li-Ya and Li, Hui and Feng, Liang and Guan, Ying-Shi and Chen, Yu-Zhe and Duan, Chun-Feng and Wu, Li-Zhu and Guan, Ya-Feng and Tung, Chen-Ho and Yang, Qing-Zheng},
  title     = {Bodipy-based Fluorometric Sensor Array for the Highly Sensitive Identification of Heavy-metal Ions},
  journal   = {Anal. Chim. Acta},
  year      = {2013},
  volume    = {775},
  pages     = {93--99},
  doi       = {10.1016/j.aca.2013.03.013},
  publisher = {Elsevier},
}

@Article{Kowada2015,
  author    = {Kowada, Toshiyuki and Maeda, Hiroki and Kikuchi, Kazuya},
  title     = {Bodipy-based Probes for the Fluorescence Imaging of Biomolecules in Living Cells},
  journal   = {Chem. Soc. Rev.},
  year      = {2015},
  volume    = {44},
  number    = {14},
  pages     = {4953--4972},
  doi       = {10.1039/C5CS00030K},
  publisher = {Royal Society of Chemistry},
}

@Article{Zhao2019,
  author  = {Kandrashkin, Y. E. and Wang, Z. and Sukhanov, A. A. and Hou, Y. and Zhang, X. and Liu, Y. and Voronkova, V. K. and Zhao, J.},
  title   = {Balance between Triplet States in Photoexcited Orthogonal BODIPY Dimers},
  journal = {J. Phys. Chem. Lett.},
  year    = {2019},
  volume  = {10},
  pages   = {4157},
  doi     = {10.1021/acs.jpclett.9b01741},
}

@Article{Kandrashkin2024,
	author  = {Wu, Yanran and Cao, Huaiman and Bakirov, Marcel M. and Sukhanov, Andrey A. and Li, Jiayu and Liao, Sheng and Xiao, Xiao and Zhao, Jianzhang and Li, Ming-De and Kandrashkin, Yuri E.},
	title   = {A Rational Way to Control the Triplet State Wave Function Confinement of Organic Chromophores: Effect of the Connection Sites and Spin Density Distribution-Guided Molecular Structure Design Principles in BODIPY Dimers},
	journal = {J. Phys. Chem. Lett.},
	year    = {2024},
	volume  = {15},
	number  = {4},
	pages   = {959-968},
	doi     = {10.1021/acs.jpclett.3c03225},
}

@Article{Montero2018,
  author    = {Montero, Ra{\'u}l and Mart{\'\i}nez-Mart{\'\i}nez, Virginia and Longarte, Asier and Epelde-Elezcano, Nerea and Palao, Eduardo and Lamas, Iker and Manzano, Hegoi and Agarrabeitia, Antonia R. and L{\'o}pez-Arbeloa, {\'I}{\~n}igo and Ortiz, Mar{\'\i}a J. and Garcia-Moreno, Inmaculada},
  title     = {Singlet Fission Mediated Photophysics of BODIPY Dimers},
  journal   = {J. Phys. Chem. Lett.},
  year      = {2018},
  volume    = {9},
  number    = {3},
  pages     = {641--646},
  doi       = {10.1021/acs.jpclett.7b03074},
  publisher = {ACS Publications},
}

@Article{Casanova2022,
  author    = {Garc{\'\i}a-Moreno, Inmaculada and Postils, Ver{\`o}nica and Rebollar, Esther and Ortiz, Maria J and Agarrabeitia, Antonia R and Casanova, David},
  title     = {Generation of Multiple Triplet States in an Orthogonal BODIPY Dimer: A Breakthrough Spectroscopic and Theoretical Approach},
  journal   = {Phys. Chem. Chem. Phys.},
  year      = {2022},
  volume    = {24},
  number    = {10},
  pages     = {5929--5938},
  doi       = {10.1039/d1cp05730h},
  publisher = {Royal Society of Chemistry},
}

@Article{Loudet2007,
  author    = {Loudet, Aurore and Burgess, Kevin},
  title     = {Bodipy Dyes and Their Derivatives: Syntheses and Spectroscopic Properties},
  journal   = {Chem Rev},
  year      = {2007},
  volume    = {107},
  number    = {11},
  pages     = {4891--4932},
  doi       = {10.1021/cr078381n},
  publisher = {ACS Publications},
}

@Article{Becke1993,
  author    = {Axel D. Becke},
  title     = {Density-functional Thermochemistry. {III}. the Role of Exact Exchange},
  journal   = {J. Chem. Phys.},
  year      = {1993},
  volume    = {98},
  number    = {7},
  pages     = {5648-5652},
  month     = {apr},
  doi       = {10.1063/1.464913},
  publisher = {{AIP} Publishing},
}

@Article{Granovsky2011,
  author    = {Granovsky, Alexander A},
  title     = {Extended Multi-configuration Quasi-degenerate Perturbation Theory: The New Approach to Multi-state Multi-reference Perturbation Theory},
  journal   = {J. Chem. Phys.},
  year      = {2011},
  volume    = {134},
  number    = {21},
  pages     = {214113},
  month     = jun,
  issn      = {1089-7690},
  doi       = {10.1063/1.3596699},
  publisher = {American Institute of Physics},
}

@InCollection{Dunning2013,
	author    = {Dunning, T. H. and Hay, P. J.},
	title     = {Gaussian Basis Sets for Molecular Calculations},
	booktitle = {Methods of Electronic Structure Theory},
	publisher = {Springer US},
	year      = {2013},
	editor    = {Schaefer III, H. F.},
	volume    = {3},
	series    = {Modern Theoretical Chemistry},
	chapter   = {1},
	pages     = {1-27},
	address   = {New York},
	doi       = {10.1007/978-1-4757-0887-5_1},
}

@Article{Gordon2020,
  author    = {Barca, Giuseppe M. J. and Bertoni, Colleen and Carrington, Laura and Datta, Dipayan and De Silva, Nuwan and Deustua, J. Emiliano and Fedorov, Dmitri G. and Gour, Jeffrey R. and Gunina, Anastasia O. and Guidez, Emilie and Harville, Taylor and Irle, Stephan and Ivanic, Joe and Kowalski, Karol and Leang, Sarom S. and Li, Hui and Li, Wei and Lutz, Jesse J. and Magoulas, Ilias and Mato, Joani and Mironov, Vladimir and Nakata, Hiroya and Pham, Buu Q. and Piecuch, Piotr and Poole, David and Pruitt, Spencer R. and Rendell, Alistair P. and Roskop, Luke B. and Ruedenberg, Klaus and Sattasathuchana, Tosaporn and Schmidt, Michael W. and Shen, Jun and Slipchenko, Lyudmila and Sosonkina, Masha and Sundriyal, Vaibhav and Tiwari, Ananta and Galvez Vallejo, Jorge L. and Westheimer, Bryce and W{\l}och, Marta and Xu, Peng and Zahariev, Federico and Gordon, Mark S.},
  title     = {Recent Developments in the General Atomic and Molecular Electronic Structure System},
  journal   = {J. Chem. Phys.},
  year      = {2020},
  volume    = {152},
  number    = {15},
  pages     = {154102},
  doi       = {10.1063/5.0005188},
  publisher = {AIP Publishing},
}

@Article{Casanova2021,
  author    = {Postils, Ver{\`o}nica and Ruip{\'e}rez, Fernando and Casanova, David},
  title     = {Mild Open-shell Character of BODIPY and Its Impact on Singlet and Triplet Excitation Energies},
  journal   = {J. Chem. Theory Comput.},
  year      = {2021},
  volume    = {17},
  number    = {9},
  pages     = {5825--5838},
  doi       = {10.1021/acs.jctc.1c00544.s001},
  publisher = {ACS Publications},
}

@Article{Michl2018,
  author  = {Wen, Jin and Han, Bowen and Havlas, Zdeněk and Michl, Josef},
  title   = {An MS-CASPT2 Calculation of the Excited Electronic States of an Axial Difluoroborondipyrromethene (bodipy) Dimer},
  journal = {J. Chem. Theory Comput.},
  year    = {2018},
  volume  = {14},
  number  = {8},
  pages   = {4291-4297},
  doi     = {10.1021/acs.jctc.8b00136},
  eprint  = {https://doi.org/10.1021/acs.jctc.8b00136},
}

@Article{Zirzlmeier2016,
  author    = {Zirzlmeier, Johannes and Casillas, Rub{\'e}n and Reddy, S Rajagopala and Coto, Pedro B and Lehnherr, Dan and Chernick, Erin T and Papadopoulos, Ilias and Thoss, Michael and Tykwinski, Rik R and Guldi, Dirk M},
  title     = {Solution-based Intramolecular Singlet Fission in Cross-conjugated Pentacene Dimers},
  journal   = {Nanoscale},
  year      = {2016},
  volume    = {8},
  number    = {19},
  pages     = {10113--10123},
  doi       = {10.1039/c6nr02493a},
  publisher = {Royal Society of Chemistry},
}

@Article{Basel2017,
	author    = {Basel, Bettina S. and Zirzlmeier, Johannes and Hetzer, Constantin and Phelan, Brian T. and Krzyaniak, Matthew D. and Reddy, S. Rajagopala and Coto, Pedro B. and Horwitz, Noah E. and Young, Ryan M. and White, Fraser J. and Hampel, Frank and Clark, Timothy and Thoss, Michael and Tykwinski, Rik R. and Wasielewski, Michael R. and Guldi, Dirk M.},
	title     = {Unified Model for Singlet Fission within a Non-conjugated Covalent Pentacene Dimer},
	journal   = {Nat. Commun.},
	year      = {2017},
	volume    = {8},
	pages     = {15171},
	month     = {may},
	doi       = {10.1038/ncomms15171},
	publisher = {Springer Nature},
}

@Article{Reddy2018,
  author    = {Reddy, S Rajagopala and Coto, Pedro B and Thoss, Michael},
  title     = {Intramolecular Singlet Fission: Insights from Quantum Dynamical Simulations},
  journal   = {J. Phys. Chem. Lett.},
  year      = {2018},
  volume    = {9},
  number    = {20},
  pages     = {5979--5986},
  month     = sep,
  issn      = {1948-7185},
  doi       = {10.1021/acs.jpclett.8b02674},
  publisher = {ACS Publications},
}

@Article{Reddy2019,
  author    = {Reddy, S Rajagopala and Coto, Pedro B and Thoss, Michael},
  title     = {Quantum Dynamical Simulation of Intramolecular Singlet Fission in Covalently Coupled Pentacene Dimers},
  journal   = {J. Chem. Phys.},
  year      = {2019},
  volume    = {151},
  number    = {4},
  pages     = {044307},
  month     = jul,
  issn      = {1089-7690},
  doi       = {10.1063/1.5109897},
  publisher = {AIP Publishing LLC},
}

@Article{Nakamura2001,
  author    = {Nakamura, Hisao and Truhlar, Donald G},
  title     = {The Direct Calculation of Diabatic States Based on Configurational Uniformity},
  journal   = {J. Phys. Chem. Lett.},
  year      = {2001},
  volume    = {115},
  number    = {22},
  pages     = {10353--10372},
  doi       = {10.1063/1.1412879},
  publisher = {American Institute of Physics},
}

@Article{Nakamura2002,
  author    = {Nakamura, Hisao and Truhlar, Donald G},
  title     = {Direct Diabatization of Electronic States by the Fourfold Way. Ii. Dynamical Correlation and Rearrangement Processes},
  journal   = {J. Chem. Phys.},
  year      = {2002},
  volume    = {117},
  number    = {12},
  pages     = {5576--5593},
  doi       = {10.1063/1.1500734},
  publisher = {American Institute of Physics},
}

@Article{Berkelbach2013,
  author    = {Berkelbach, Timothy C. and Hybertsen, Mark S. and Reichman, David R.},
  title     = {Microscopic Theory of Singlet Exciton Fission. Ii. Application to Pentacene Dimers and the Role of Superexchange},
  journal   = {J. Chem. Phys.},
  year      = {2013},
  volume    = {138},
  number    = {11},
  pages     = {114103},
  issn      = {0021-9606},
  doi       = {10.1063/1.4794427},
  publisher = {American Institute of Physics (AIP)},
}

@Article{Reddy2018a,
	author   = {S. Rajagopala Reddy and Pedro B. Coto and Michael Thoss},
	title    = {Theoretical Study of Intramolecular Singlet Fission in Xanthene-bonded Pentacene Dimers},
	journal  = {Chem. Phys.},
	year     = {2018},
	volume   = {515},
	pages    = {628 - 634},
	issn     = {0301-0104},
	doi      = {10.1016/j.chemphys.2018.07.027},
}

@Book{Nitzan2006,
	Title                    = {Chemical Dynamics in Condensed Phases: Relaxation, Transfer and Reactions in Condensed Molecular Systems},
	Author                   = {Nitzan, A.},
	Publisher                = {OUP Oxford},
	Year                     = {2006}
}

@Article{Sebastian1981,
  author    = {Sebastian, L and Weiser, G and B{\"a}ssler, H},
  title     = {Charge Transfer Transitions in Solid Tetracene and Pentacene Studied by Electroabsorption},
  journal   = {Chem. Phys.},
  year      = {1981},
  volume    = {61},
  number    = {1-2},
  pages     = {125--135},
  doi       = {10.1016/0301-0104(81)85055-0},
  publisher = {Elsevier},
}

@Article{Goyal2024,
  author    = {Goyal, Sophiya and Reddy, S Rajagopala},
  title     = {Investigation of Excited States of BODIPY Derivatives and Non-orthogonal Dimers in the Perspective of Singlet Fission},
  journal   = {Phys. Chem. Chem. Phys.},
  year      = {2024},
  volume    = {26},
  pages     = {26398-26408},
  doi       = {10.1039/d4cp02656j},
  publisher = {Royal Society of Chemistry},
}

@Article{Tian2021,
  author    = {Tian, Dandan and Pan, Hongfei and Zhang, Yongjie and Ren, Xiang-Kui and Chen, Zhijian},
  title     = {NIR Absorbing Dimeric Aza-BODIPY Dye with J-type Aggregation and Photothermal Properties},
  journal   = {Tetrahedron Lett.},
  year      = {2021},
  volume    = {76},
  pages     = {153216},
  doi       = {10.1016/j.tetlet.2021.153216},
  publisher = {Elsevier},
}

@misc{Frisch2009,
  title        = {Gaussian 09, Revision E.01},
  author       = {Frisch, M. J. and Trucks, G. W. and Schlegel, H. B. and Scuseria, G. E. and Robb, M. A. and Cheeseman, J. R. and Scalmani, G. and Barone, V. and Petersson, G. A. and Nakatsuji, H. and Li, X. and Caricato, M. and Marenich, A. and Bloino, J. and Janesko, B. G. and Gomperts, R. and Mennucci, B. and Hratchian, H. P. and Ortiz, J. V. and Izmaylov, A. F. and Sonnenberg, J. L. and Williams-Young, D. and Ding, F. and Lipparini, F. and Egidi, F. and Goings, J. and Peng, B. and Petrone, A. and Henderson, T. and Ranasinghe, D. and Zakrzewski, V. G. and Gao, J. and Rega, N. and Zheng, G. and Liang, W. and Hada, M. and Ehara, M. and Toyota, K. and Fukuda, R. and Hasegawa, J. and Ishida, M. and Nakajima, T. and Honda, Y. and Kitao, O. and Nakai, H. and Vreven, T. and Throssell, K. and Montgomery, J. A., Jr. and Peralta, J. E. and Ogliaro, F. and Bearpark, M. and Heyd, J. J. and Brothers, E. and Kudin, K. N. and Staroverov, V. N. and Keith, T. and Kobayashi, R. and Normand, J. and Raghavachari, K. and Rendell, A. and Burant, J. C. and Iyengar, S. S. and Tomasi, J. and Cossi, M. and Millam, J. M. and Klene, M. and Adamo, C. and Cammi, R. and Ochterski, J. W. and Martin, R. L. and Morokuma, K. and Farkas, O. and Foresman, J. B. and Fox, D. J.},
  year         = {2016},
  organization = {Gaussian, Inc.},
  address      = {Wallingford, CT},
}

@Article{Jr1989,
  author    = {Dunning Jr, Thom H},
  title     = {Gaussian Basis Sets for Use in Correlated Molecular Calculations. I. the Atoms Boron through Neon and Hydrogen},
  journal   = {J. Chem. Phys.},
  year      = {1989},
  volume    = {90},
  number    = {2},
  pages     = {1007--1023},
  doi       = {10.1063/1.456153},
  publisher = {American Institute of Physics},
}

@Article{Guo2022,
  author    = {Guo, Xing and Yang, Jinming and Li, Mao and Zhang, Fan and Bu, Weibin and Li, Heng and Wu, Qinghua and Yin, Dengke and Jiao, Lijuan and Hao, Erhong},
  title     = {Unique Double Intramolecular and Intermolecular Exciton Coupling in Ethene-bridged Aza-BODIPY Dimers for High-efficiency Near-infrared Photothermal Conversion and Therapy},
  journal   = {Angew. Chem. Int. Ed.},
  year      = {2022},
  volume    = {61},
  number    = {44},
  pages     = {e202211081},
  publisher = {Wiley Online Library},
}

@Article{Batat2011,
  author    = {Batat, Pinar and Cantuel, Martine and Jonusauskas, Gediminas and Scarpantonio, Luca and Palma, Aniello and O’Shea, Donal F and McClenaghan, Nathan D},
  title     = {BF2-azadipyrromethenes: Probing the Excited-state Dynamics of a NIR Fluorophore and Photodynamic Therapy Agent},
  journal   = {J. Phys. Chem. A},
  year      = {2011},
  volume    = {115},
  number    = {48},
  pages     = {14034--14039},
  publisher = {ACS Publications},
}

@Article{Awuah2012,
  author    = {Awuah, Samuel G and You, Youngjae},
  title     = {Boron Dipyrromethene (bodipy)-based Photosensitizers for Photodynamic Therapy},
  journal   = {RSC Adv},
  year      = {2012},
  volume    = {2},
  number    = {30},
  pages     = {11169--11183},
  publisher = {Royal Society of Chemistry},
}

@Article{Killoran2002,
  author    = {Killoran, John and Allen, Lorcan and Gallagher, John F and Gallagher, William M and O'Shea, Donal F},
  title     = {Synthesis of BF2 Chelates of Tetraarylazadipyrromethenes and Evidence for Their Photodynamic Therapeutic Behaviour},
  journal   = {Chem. Commun.},
  year      = {2002},
  number    = {17},
  pages     = {1862--1863},
  publisher = {Royal Society of Chemistry},
}

@Article{Ni2014,
  author    = {Ni, Yong and Wu, Jishan},
  title     = {Far-red and near Infrared Bodipy Dyes: Synthesis and Applications for Fluorescent Ph Probes and Bio-imaging},
  journal   = {Org. Biomol. Chem.},
  year      = {2014},
  volume    = {12},
  number    = {23},
  pages     = {3774--3791},
  doi       = {10.1039/C3OB42554A},
  publisher = {Royal Society of Chemistry},
}

@Article{Wael1977,
  author   = {de Wael, E. Vos and Pardoen, J. A. and van Koeveringe, J. A. and Lugtenburg, J.},
  journal  = {Recl. Trav. Chim. Pays-Bas},
  title    = {Pyrromethene-BF2 complexes (4,\`{4}-difluoro-4-bora-3a,4a-diaza-s-indacenes). Synthesis and luminescence properties},
  year     = {1977},
  number   = {12},
  pages    = {306-309},
  volume   = {96},
  doi      = {https://doi.org/10.1002/recl.19770961205},
  eprint   = {https://onlinelibrary.wiley.com/doi/pdf/10.1002/recl.19770961205},
  url      = {https://onlinelibrary.wiley.com/doi/abs/10.1002/recl.19770961205},
}

@Article{Goyal2026,
  author    = {Sophiya Goyal and S. Rajagopala Reddy},
  journal   = {J. Phys. Chem. A},
  title     = {Exploring Multiexciton Generation in an Asymmetric Aza-BODIPY Dimer},
  year      = {2026},
  issn      = {1089-5639},
  number    = {9},
  pages     = {1767-1779},
  volume    = {130},
  doi       = {10.1021/acs.jpca.5c06838},
  publisher = {American Chemical Society (ACS)},
}

@Article{Majumder2024,
  author    = {Kanad Majumder and Soham Mukherjee and Jungjin Park and Woojae Kim and Andrew J. Musser and Satish Patil},
  journal   = {Angewandte Chemie},
  title     = {The Acetylene Bridge in Intramolecular Singlet Fission: A Boon or A Nuisance?},
  year      = {2024},
  issn      = {0044-8249},
  number    = {52},
  volume    = {136},
  doi       = {10.1002/ange.202408615},
  publisher = {Wiley},
}

@article{Goyal_arxiv2026,
  author  = {Sophiya Goyal and S. Rajagopala Reddy},
  title   = {Regio-Connectivity and Torsional Angle Effects on Singlet Fission and SOCT-ISC in Aza-BODIPY Dimers},
  journal = {arXiv preprint arXiv:2604.03011,},
  year    = {2026}
}

@Article{Zirzlmeier2015,
  author    = {Zirzlmeier, Johannes and Lehnherr, Dan and Coto, Pedro B. and Chernick, Erin T. and Casillas, Rubén and Basel, Bettina S. and Thoss, Michael and Tykwinski, Rik R. and Guldi, Dirk M.},
  journal   = {Proc. Natl. Acad. Sci.},
  title     = {Singlet fission in pentacene dimers},
  year      = {2015},
  issn      = {1091-6490},
  month     = Apr,
  number    = {17},
  pages     = {5325--5330},
  volume    = {112},
  doi       = {10.1073/pnas.1422436112},
  publisher = {Proceedings of the National Academy of Sciences},
}

@Article{Basel2018,
  author    = {Bettina S. Basel and Johannes Zirzlmeier and Constantin Hetzer and S. Rajagopala Reddy and Brian T. Phelan and Matthew D. Krzyaniak and Michel K. Volland and Pedro B. Coto and Ryan M. Young and Timothy Clark and Michael Thoss and Rik R. Tykwinski and Michael R. Wasielewski and Dirk M. Guldi},
  journal   = {Chem},
  title     = {Evidence for Charge-Transfer Mediation in the Primary Events of Singlet Fission in a Weakly Coupled Pentacene Dimer},
  year      = {2018},
  issn      = {2451-9294},
  number    = {5},
  pages     = {1092-1111},
  volume    = {4},
  doi       = {10.1016/j.chempr.2018.04.006},
  publisher = {Elsevier BV},
}

\clearpage
\section*{Appendix}

\begin{table}[H]
	\caption[Excited-state properties of dimer {[}1{]} computed at SA15- and SA8-XMCQDPT levels.]{The excitation energy (in eV units), oscillator strength (in dimensionless) and dipole moment (in Debye units) and electronic configuration of the dimer [1] calculated by using SA15-XMCQDPT(8,8) and SA8-XMCQDPT(4,4) level of theory.}
	\small
	\centering
\begin{tabular}{cccccccc}
		\toprule
		\multirow{2}{*}{S. No.} & \multicolumn{4}{c}{SA15-XPT(8,8)} & \multicolumn{2}{c}{SA8-XPT(4,4)} & \multirow{2}{*}{Char.$^a$}   \\
		& $\Delta E$ & $f$ & $\mu$ & config$^b$ & $\Delta E$ & config$^b$  &  \\
		\toprule	
		S$_\text{0}$ & 0.000 &    0.000   & 2.45 & 22220000 & 0.000 & 2200 &   GS     \\
		\midrule
		\multirow{2}{*}{S$_\text{1}$} & \multirow{2}{*}{1.950} & \multirow{2}{*}{1.729}  & \multirow{2}{*}{0.94} & 222+-000 & \multirow{2}{*}{1.964} & 2+-0 &  \multirow{2}{*}{LE$_1$}      \\
		&       &       &      & 22+20-00 &       & +20- &        \\
		\midrule
		\multirow{5}{*}{S$_\text{2}$} & \multirow{5}{*}{2.257} & \multirow{5}{*}{$<$0.001}  & \multirow{5}{*}{2.16} & 22+2-000 & \multirow{5}{*}{2.298} & +2-0 & \multirow{5}{*}{ME}       \\
		&       &       &      & 22202000 &       & 2020 &        \\
		&       &       &      & 22++- -00 &       & ++- - &        \\
\bottomrule		
	\end{tabular}\\
	\small $^a$In localized orbitals. $^b$In delocalized orbitals; the description was obtained using the same procedure as in Figure S3 of ref.~\cite{Goyal2026}.  
	\label{tab:ortho}
\end{table}

\begin{table}[H]
	\caption {Excited-state parameters for [2] (cf. Table~\ref{tab:ortho}).}
	\small
	\centering
	\begin{tabular}{cccccccc}
		\toprule
		\multirow{2}{*}{S. No.} & \multicolumn{4}{c}{SA15-XPT(8,8)} & \multicolumn{2}{c}{SA8-XPT(4,4)} & \multirow{2}{*}{Char.$^a$}   \\
		& $\Delta E$ & $f$ & $\mu$ & config$^b$ & $\Delta E$ & config$^b$  &  \\
		\toprule	
		S$_\text{0}$ & 0.000 &  0.000     & 3.43 & 22220000 &  0.000 & 2200 &    GS    \\
		\midrule
		\multirow{2}{*}{S$_\text{1}$} & \multirow{2}{*}{2.081} & \multirow{2}{*}{2.089}  & \multirow{2}{*}{3.77} & 222+0-00 &  \multirow{2}{*}{2.321} & 2+0-  &   \multirow{2}{*}{LE$_1$}      \\
		&       &       &      & 22+2-000 &        & +2-0  &        \\
		\midrule 
		\multirow{2}{*}{S$_\text{2}$} & \multirow{2}{*}{2.322} & \multirow{2}{*}{0.393}  & \multirow{2}{*}{3.94} & 222+-000 & \multirow{2}{*}{2.050} & 2+-0  &     \multirow{2}{*}{LE$_1$}    \\
		&       &       &      & 22+20-00 &       & +20-  &        \\
		\midrule 
		\multirow{5}{*}{S$_\text{3}$} & \multirow{5}{*}{2.447} & \multirow{5}{*}{$<$0.001}  & \multirow{5}{*}{2.67} & 22++- -00 &  \multirow{5}{*}{2.478} & ++- -  &   \multirow{5}{*}{ME}      \\
\bottomrule		
	\end{tabular}\\
	\small $^a$In localized orbitals. $^b$In delocalized orbitals; the description was obtained using the same procedure as in Figure S3 of ref.~\cite{Goyal2026}.
	\label{tab:meta}
\end{table}

\begin{table}[H]
	\caption{Excited-state parameters for [3] (cf. Table~\ref{tab:ortho}).}
	\small
	\centering
	\begin{tabular}{cccccccc}
		\toprule
		\multirow{2}{*}{S. No.} & \multicolumn{4}{c}{SA15-XPT(8,8)} &  \multicolumn{2}{c}{SA8-XPT(4,4)} & \multirow{2}{*}{Char.$^a$}   \\
		& $\Delta E$ & $f$ & $\mu$ & config$^b$ & $\Delta E$ & config$^b$  &  \\
		\toprule	
		S$_\text{0}$ & 0.000 &   0.000    & 0.30 & 22220000 & 0.000 &  2200  &  GS      \\
		\midrule
		\multirow{2}{*}{S$_\text{1}$} & \multirow{2}{*}{1.976} & \multirow{2}{*}{2.054}  & \multirow{2}{*}{1.49} & 222+-000 & \multirow{2}{*}{1.929} & 2+-0  &   \multirow{2}{*}{LE$_1$}      \\
		&       &        &      & 22+20-00 &       & +20-  &        \\
		\midrule
		\multirow{6}{*}{S$_\text{2}$} & \multirow{6}{*}{2.170} & \multirow{6}{*}{$<$0.001}  & \multirow{6}{*}{0.21} & 22202000 & \multirow{6}{*}{2.190} & 2020  &   \multirow{6}{*}{ME}      \\
		&       &        &      & 222+0-00 &       & 2+0-   &        \\
		&       &        &      & 22+2-000 &       & +2-0  &        \\
\bottomrule		
	\end{tabular}\\
	\small $^a$In localized orbitals. $^b$In delocalized orbitals; the description was obtained using the same procedure as in Figure S3 of ref.~\cite{Goyal2026}. 
	\label{tab:para}
\end{table}

\begin{table}[H]
	\caption{Excited-state parameters for [4] (cf. Table~\ref{tab:ortho}).}
	\small
	\centering
	\begin{tabular}{cccccccc}
		\toprule
		\multirow{2}{*}{S. No.} & \multicolumn{4}{c}{SA15-XPT(8,8)} &  \multicolumn{2}{c}{SA8-XPT(4,4)} & \multirow{2}{*}{Char.$^a$} \\
		& $\Delta E$ & $f$ & $\mu$ & config$^b$  & $\Delta E$ & config$^b$ & \\
		\toprule
		S$_\text{0}$ & 0.000 &   0.000   & 2.43 & 22220000  &  &   & GS \\
		\midrule
		\multirow{2}{*}{S$_\text{1}$} & \multirow{2}{*}{2.069} & \multirow{2}{*}{2.463} & \multirow{2}{*}{2.14} & 222+-000  & \multirow{2}{*}{2.009}   &2+-0    &\multirow{2}{*}{LE$_1$} \\
		&       &      &      & 22+20-00    &       &   +20-    & \\
		\midrule
		\multirow{6}{*}{S$_\text{2}$} & \multirow{6}{*}{2.242} & \multirow{6}{*}{$<$0.001} & \multirow{6}{*}{2.19} & 22202000     & \multirow{6}{*}{2.234} & 2020    &\multirow{6}{*}{ME} \\
		&       &      &      & 22++- -00   &       & ++- -    & \\
		&       &      &      & 22200200     &       & 2002    & \\
\bottomrule
	\end{tabular}\\
	\small $^a$In localized orbitals. $^b$In delocalized orbitals; the description was obtained using the same procedure as in Figure S3 of ref.~\cite{Goyal2026}.
	\label{tab:acet-ortho}
\end{table}

\begin{table}[H]
	\caption{Excited-state parameters for [5] (cf. Table~\ref{tab:ortho}).}	
	\small
	\centering
	\begin{tabular}{cccccccc}
		\toprule
		\multirow{2}{*}{S. No.} & \multicolumn{4}{c}{SA15-XPT(8,8)}  & \multicolumn{2}{c}{SA8-XPT(4,4)} & \multirow{2}{*}{Char.$^a$} \\
		& $\Delta E$ & $f$ & $\mu$ & config$^b$  & $\Delta E$ & config$^b$ & \\
		\toprule
		S$_\text{0}$ & 0.000 &   0.000    & 1.01 & 22220000       & 0.000 & 2200    & GS \\
		\midrule
		\multirow{2}{*}{S$_\text{1}$} & \multirow{2}{*}{2.053} & \multirow{2}{*}{2.214}  & \multirow{2}{*}{1.61} & 222+-000   & \multirow{2}{*}{2.087} & 2+-0    & \multirow{2}{*}{LE$_1$}\\
		&       &       &      & 22+20-00    &           & +20-    & \\
		\midrule
		\multirow{4}{*}{S$_\text{2}$} & \multirow{4}{*}{2.243} & \multirow{4}{*}{0.521}  & \multirow{4}{*}{1.62} & 222+0-00   & \multirow{4}{*}{2.248} & 2+0-    & \multirow{4}{*}{LE$_1$} \\
		&       &       &      & 22+2-000     &       & +2-0    & \\
		&       &       &      & 220+2-00     &       & +0-2    & \\
\bottomrule
	\end{tabular}\\
	\label{tab:acet-meta}
	\small $^a$In localized orbitals. $^b$In delocalized orbitals; the description was obtained using the same procedure as in Figure S3 of ref.~\cite{Goyal2026}.
\end{table}

\begin{table}[H]
	\caption{Excited-state parameters for [6] (cf. Table~\ref{tab:ortho}).}
	\small
	\centering
	\begin{tabular}{cccccccc}
		\toprule
		\multirow{2}{*}{S. No.} & \multicolumn{4}{c}{SA15-XPT(8,8)}  & \multicolumn{2}{c}{SA8-XPT(4,4)} & \multirow{2}{*}{Char.$^a$} \\
		& $\Delta E$ & $f$ & $\mu$ & config$^b$ & $\Delta E$ & config$^b$ & \\
		\toprule
		S$_\text{0}$ & 0.000 &    0.000   & 1.98 & 22220000     & 0.000 & 2200    &GS\\
		\midrule
		\multirow{2}{*}{S$_\text{1}$} & \multirow{2}{*}{1.945} & \multirow{2}{*}{2.546}  & \multirow{2}{*}{2.13} & 222+-000    & \multirow{2}{*}{2.012} & 2+-0    & \multirow{2}{*}{LE$_1$} \\
		&       &       &      & 22+20-00       &       & +20-    & \\
		\midrule
		\multirow{5}{*}{S$_\text{2}$} & \multirow{5}{*}{2.155} & \multirow{5}{*}{$<$0.001}  & \multirow{5}{*}{1.94} & 22202000   & \multirow{5}{*}{2.195} & 2020    & \multirow{5}{*}{ME} \\
		&       &       &      & 22++- -00     &       & ++- -    & \\
		&       &       &      & 22200200     &       & 2002    & \\
\bottomrule
	\end{tabular}\\
	\small $^a$In localized orbitals. $^b$In delocalized orbitals; the description was obtained using the same procedure as in Figure S3 of ref.~\cite{Goyal2026}.
	\label{tab:acet-para}
\end{table}

\begin{table}[H]
	\caption{Excited-state parameters for [7] (cf. Table~\ref{tab:ortho}).}
	\small
	\centering
	\begin{tabular}{cccccccc}
		\toprule
		\multirow{2}{*}{S. No.} & \multicolumn{4}{c}{SA15-XPT(8,8)}  & \multicolumn{2}{c}{SA8-XPT(4,4)} & \multirow{2}{*}{Char.$^a$}   \\
		& $\Delta E$ & $f$ & $\mu$ & config$^b$  & $\Delta E$ & config$^b$  &  \\
		\toprule	
		S$_\text{0}$ & 0.000 & 0.000 & 2.16 & 22220000 & 0.000 &  2200      &    GS    \\
		\midrule
		\multirow{2}{*}{S$_\text{1}$} & \multirow{2}{*}{1.929} & \multirow{2}{*}{0.577}  & \multirow{2}{*}{4.81} & 222+0-00    & \multirow{2}{*}{2.082} & 2+-0       &   \multirow{2}{*}{LE$_1$}    \\
		&       &       &      & 222+-000  &       & +2-0  & \\
		\midrule          
		\multirow{2}{*}{S$_\text{2}$} & \multirow{2}{*}{1.978} & \multirow{2}{*}{1.951}  & \multirow{2}{*}{6.95} &  222+-000   & \multirow{2}{*}{1.955} &  2+-0     & \multirow{2}{*}{LE$_1$}       \\
		&       &   &  & 222+0-00     &       & +20- & \\
		
		\midrule
		
		\multirow{3}{*}{S$_\text{3}$} & \multirow{3}{*}{2.322} & \multirow{3}{*}{$<$0.001} & \multirow{3}{*}{2.08} & 22++- -00              & \multirow{3}{*}{2.212} & ++- -      &   \multirow{4}{*}{ME}    \\
\bottomrule		
	\end{tabular}\\
	\small $^a$In localized orbitals. $^b$In delocalized orbitals; the description was obtained using the same procedure as in Figure S3 of ref.~\cite{Goyal2026}.
	\label{tab:xc2}
\end{table}

\begin{table}[H]
	\caption{Excited-state parameters for [8] (cf. Table~\ref{tab:ortho}).}
	\small
	\centering
	\begin{tabular}{cccccccc}
		\toprule
		\multirow{2}{*}{S. No.} & \multicolumn{4}{c}{SA15-XPT(8,8)} & \multicolumn{2}{c}{SA8-XPT(4,4)} & \multirow{2}{*}{Char.$^a$}   \\
		& $\Delta E$ & $f$ & $\mu$ & config$^b$ & $\Delta E$ & config$^b$  &  \\
		\toprule	
		S$_\text{0}$ & 0.000 & 0.000  & 88.85 & 22220000 & 0.000 & 2200  &   GS     \\
		\midrule
		\multirow{2}{*}{S$_\text{1}$} & \multirow{2}{*}{1.902} &  \multirow{2}{*}{0.896} & \multirow{2}{*}{81.93} & 222+-000 & \multirow{2}{*}{1.790} & 2+-0  &    \multirow{2}{*}{LE$_1$}   \\
		&       &       &      & 22+20-00 &       & +20-& \\
		
		\midrule
		\multirow{2}{*}{S$_\text{2}$} & \multirow{2}{*}{2.256} &  \multirow{2}{*}{0.016} & \multirow{2}{*}{88.03} & 222+0-00   & \multirow{2}{*}{2.024} &  2+0- &    \multirow{2}{*}{LE$_1$}    \\
		&       &       &      &22+2-000&       &+2-0 & \\
		
		\midrule  
		\multirow{5}{*}{S$_\text{3}$} & \multirow{5}{*}{2.316} & \multirow{5}{*}{0.033}  &  \multirow{5}{*}{77.45} & 22202000 & \multirow{5}{*}{2.260} & 2020&     \multirow{5}{*}{ME}   \\
\bottomrule		
	\end{tabular}\\
	\small $^a$In localized orbitals. $^b$In delocalized orbitals; the description was obtained using the same procedure as in Figure S3 of ref.~\cite{Goyal2026}.
	\label{tab:xc2-without-acet}
\end{table}

\begin{table}[H]
	\caption{Excited-state parameters for [9] (cf. Table~\ref{tab:ortho}).}
	\small
	\centering
	\begin{tabular}{cccccccc}
		\toprule
		\multirow{2}{*}{S. No.} & \multicolumn{4}{c}{SA15-XPT(8,8)} & \multicolumn{2}{c}{SA8-XPT(4,4)} & \multirow{2}{*}{Char.$^a$}   \\
		& $\Delta E$ & $f$ & $\mu$ & config$^b$ &  $\Delta E$ & config$^b$  &  \\
		\toprule	
		S$_\text{0}$ & 0.000 & 0.000  & 15.59 & 22220000        & 0.000 & 2200       &     GS   \\
		\midrule
		\multirow{4}{*}{S$_\text{1}$} & \multirow{4}{*}{2.214} & \multirow{4}{*}{1.002}  & \multirow{4}{*}{16.24} & 222+-000        & \multirow{4}{*}{2.148} &2+-0    &    \multirow{4}{*}{LE$_1$}    \\
		&       &       &      & 22+20-00 &       &+20- & \\
		&       &       &      &  &       &2+0- & \\
		&       &       &      &  &       &+2-0 & \\
		\midrule
		\multirow{2}{*}{S$_\text{2}$} & \multirow{2}{*}{2.259} & \multirow{2}{*}{$<$0.001} & \multirow{2}{*}{9.23} & 222+-000       & \multirow{2}{*}{2.231} &  \multirow{2}{*}{2+0-}      &  \multirow{2}{*}{LE$_1$}  \\
\bottomrule		
	\end{tabular}\\
	\small $^a$In localized orbitals. $^b$In delocalized orbitals; the description was obtained using the same procedure as in Figure S3 of ref.~\cite{Goyal2026}.
	\label{tab:adamantane}
\end{table}

\end{document}